\documentclass[preprint2]{aastex}

\usepackage{amsmath,amssymb,graphicx,natbib}

\newcommand{\kepler}{\textit{Kepler}}

\newcommand{\xxii}{$\Xi$}
\newcommand{\ximax}{$\Xi_\text{max}$}

\newcommand{\singlesystems}{1481}
\newcommand{\multisystems}{155}
\newcommand{\multipairs}{262}

\shorttitle{TTV Candidates through Q6}
\shortauthors{Steffen \textit{et al.}}

\begin{document}

\title{Transit Timing Observations from \kepler: VI.  Potentially interesting candidate systems from Fourier-based statistical tests}

\author{
Jason H. Steffen\altaffilmark{1},
Eric B. Ford\altaffilmark{2},    
Jason F. Rowe\altaffilmark{8,9},    
Daniel C. Fabrycky\altaffilmark{3,4}, 
Matthew J. Holman\altaffilmark{5},  
William F. Welsh\altaffilmark{6},  
Natalie M. Batalha\altaffilmark{7}, 
William J. Borucki\altaffilmark{8}, 
Steve Bryson\altaffilmark{8},  
Douglas A. Caldwell\altaffilmark{8,9},  
David R. Ciardi\altaffilmark{10},
Jon M. Jenkins\altaffilmark{8,9},  
Hans Kjeldsen\altaffilmark{11},
David G. Koch\altaffilmark{8}, 
Andrej Pr\v sa\altaffilmark{12},
Dwight T. Sanderfer \altaffilmark{8},
Shawn Seader\altaffilmark{8},
Joseph D. Twicken\altaffilmark{8,9}, 
}
\altaffiltext{1}{Fermilab Center for Particle Astrophysics, P.O. Box 500, MS 127, Batavia, IL 60510}
\altaffiltext{2}{Astronomy Department, University of Florida, 211 Bryant Space Sciences Center, Gainesville, FL 32111, USA}
\altaffiltext{3}{UCO/Lick Observatory, University of California, Santa Cruz, CA 95064, USA}
\altaffiltext{4}{Hubble Fellow}
\altaffiltext{5}{Harvard-Smithsonian Center for Astrophysics, 60 Garden Street, Cambridge, MA 02138, USA}
\altaffiltext{6}{Astronomy Department, San Diego State University, San Diego, CA 92182-1221 , USA}
\altaffiltext{7}{San Jose State University, San Jose, CA 95192, USA}
\altaffiltext{8}{NASA Ames Research Center, Moffett Field, CA, 94035, USA}
\altaffiltext{9}{SETI Institute, Mountain View, CA, 94043, USA}
\altaffiltext{10}{NASA Exoplanet Science Institute/California Institute of Technology, Pasadena, CA 91125 USA}
\altaffiltext{11}{Department of Physics and Astronomy, Aarhus University, DK-8000 Aarhus C, Denmark}
\altaffiltext{12}{Department of Astronomy and Astrophysics, Villanova University, 800 East Lancaster Avenue, Villanova, PA 19085, USA}

\email{jsteffen@fnal.gov}

\begin{abstract}
We analyze the deviations of transit times from a linear ephemeris for the Kepler Objects of Interest (KOI) through Quarter six (Q6) of science data.  We conduct two statistical tests for all KOIs and a related statistical test for all pairs of KOIs in multi-transiting systems.  These tests identify several systems which show potentially interesting transit timing variations (TTVs).  Strong TTV systems have been valuable for the confirmation of planets and their mass measurements.  Many of the systems identified in this study should prove fruitful for detailed TTV studies.
\end{abstract}


\maketitle

\section{Introduction}

In transiting exoplanet systems, the deviations from a constant orbital period caused by planet-planet dynamical interactions have proven to be very useful in detecting and characterizing the constituent planets.  Transit timing variations (TTVs) for short-timescale and resonant interactions are particularly useful \citep{Agol:2005,Holman:2005} though secularly induced TTVs have also been studied \citep{Miralda-Escude:2002,Heyl:2007}.  The \kepler\ mission has profited greatly by the use of TTVs, having used them extensively in the confirmation of a significant fraction of its currently confirmed planets: Kepler-9 \citep{Holman:2010a}, Kepler-11 \citep{Lissauer:2011a}, Kepler-18 \citep{Cochran:2011}, and Kepler systems 23-34 \citep{Ford:2012a,Fabrycky:2012,Steffen:2012a}.  TTVs have given: important mass measurements for smaller planets that would be exceedingly difficult to obtain through radial velocity (RV) measurements; stringent limits on the presence of small planets near mean-motion resonance (MMR) with hot Jupiters \citep{Steffen:2005,Steffen:2012b} with consequent constraints on planet formation and dynamical evolution; and the identification of non-transiting planets such as, Kepler-19 \citep{Ballard:2011}---with as-yet undetermined orbital properties.

With the inital release of two quarters of data in 2011 \citep{Borucki:2011} there was an accompanying paper discussing potential TTV candidates \citep{Ford:2011}.  Now, with the release of more than a year of additional data, we revisit task of identifying systems that show interesting TTV signatures here and in a companion paper \citep{Ford:2012b} using transit times reported in \citep{Rowe:2012}.  
A number of these systems have already been identified, studied, and announced.  However, several new systems have been the subject of only cursory investigation and merit further scrutiny.  The increased time baseline of the \kepler\ data through Q6 provides many more systems with interesting TTV features as typical timescales for large TTV signals are tens of orbits---which are only now avaiable.  With continued operations the scientific yield from \kepler\ and from TTVs specifically will be particularly valuable to the study of planets as many of the orbital and physical properties of planets with small sizes and masses can only be gleaned through TTV analysis.  The extension of the \kepler\ mission, with its increased time baseline for TTV studies, will present an opportunity to learn about planets---notably including terrestrial planets and planets near the habitable zone---that will not be rivaled for many decades.

In what follows we present the results of a few straightforward statistical tests on the transit times of known Kepler Objects of Interest (KOI).  Two of these tests are applied to all individual KOIs while a third test is applied to pairs of KOIs in the same system in an effort to use correlated TTV signals as a means to improve sensitivity to dynamical interactions.  The point of these tests is not to assign rigorous absolute probabilities for the detection of a TTV signal; rather it is to identify systems that merit additional scrutiny.  The results of the tests are presented in a single, large table for all analyzed systems (Table \ref{bigoltable}).

\section{Data Reduction}

We use KOIs from the planet candidate table reported in \citep{Borucki:2011} and \citep{Batalha:2012}.  The transit times for each KOI are determined following the method described in \citep{Ford:2011} and are given in tabular form in \citep{Rowe:2012}.  Frequently with that method there are some transit times that have anomalously large deviations from a linear ephemeris or that have unusually large error bars.  In most cases these anomalies are due to some source of photometric noise or transit detections with low significance rather than from planetary dynamics.  However, in cases in which TTVs are larger than the transit duration, the algorithm may fail to find the correct solution, with the optimization algorithm not reaching the global minimum even if the individual transits are highly significant.  In those cases, we rerun the algorithm with the initial guesses close enough to the true transit times so the algorithm can model them correctly.

In order to remove non-planetary anomalous TTVs from our analysis, we reduce our timing data by requiring that the mean signal-to-noise ratio (SNR) per transit for a given KOI exceed 3.  We eliminate all transit times where the timing residuals deviate from a linear ephemeris by more than four times the Median Absolute Deviation (MAD) of all timing residuals.  We also eliminate all transit times where the timing uncertainty is larger than twice the median error of all timing uncertainties for that system.  Once this data reduction criteria has been applied, we fit a new linear ephemeris and conduct our remaining analysis on the resulting timing residuals.  Table \ref{bigoltable} shows the KOIs studied here (\singlesystems\ with 1009 that satisfy the selection criteria) along with the results of the statistical tests outlined below.

\section{Individual Objects}

The first analysis we present is applied to each individual KOI.  Here we look for a sinusoidal TTV signal with arbitrary amplitude, phase, and period.  TTV signals for systems in or near mean-motion resonance (MMR) are expected to be roughly sinusoidal as the description of their motion is dominated by a single sinusoidal component that corresponds to that MMR \citep{Agol:2005}.  While systems farther from MMR have multi-sinusoidal TTV signatures, the amplitude of the signal is significantly weaker and would not likely be marked as significant in this study \citep{Agol:2005,Steffen:2006}.  We compare this single-sinusoid model to the null model where the transit times are fit to a linear ephemeris using the standard F-ratio test with two and five parameters for the null model and the sinusoidal model respectively.  This method was first applied to a sample of hot Jupiter candidates in \citep{Steffen:2012b}.

\begin{figure}
\includegraphics[width=0.45\textwidth]{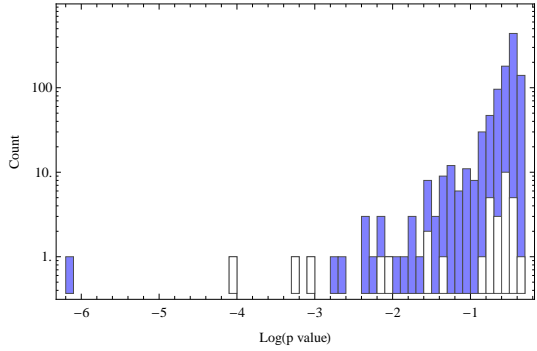}
\caption{Histogram of the p-values resulting from the F-Ratio test.  The p-value from known multiplanet systems are the light colored bars.  The two most significant systems with this test are KOI-227 and Kepler-9.\label{pvalueplot}}
\end{figure}

We do not claim that the results of this test give an accurate estimate of the statistical significance of a TTV signal, rather that this test can identify interesting systems that merit further scrutiny---as is the goal of this letter.  The results of the F-ratio test (p-values that represent the probability that the two-parameter, null model is the correct model of the two models being compared) are shown in Figure \ref{pvalueplot}.  We note that the KOIs in light colors in this and all figures are for known planets or well studied KOIs and include: 84 (Kepler-19), 137 (Kepler-18), 157 (Kepler-11), 168 (Kepler-23), 244 (Kepler-25), 250 (Kepler-26), 377 (Kepler-9), 738 (Kepler-29), 806 (Kepler-30), 841 (Kepler-27), 870 (Kepler-28), 935 (Kepler-31), 952 (Kepler-32), and 1102 (Kepler-24).

There are a few planet candidates among the group of well studied objects that have relatively small p-values, notably KOI-806 (Kepler-30) \citep{Tingley:2011,Fabrycky:2012} and KOI-377 (Kepler-9) \citep{Holman:2010a}.  However, it is noteworthy that most of the multi-transiting systems that were studied for anticorrelated TTV signatures in \citep{Ford:2012a,Fabrycky:2012,Steffen:2012a} and the planets in Kepler-11 \citep{Lissauer:2011a} do not have particularly low p-values.  This is further indication of the importance of multi-transiting planetary systems for the purposes of planet confirmation and dynamical interpretation.  A correlation analysis, as is done in \citep{Ford:2012a,Fabrycky:2012,Steffen:2012a} and in the next section of this work, can identify otherwise weak individual candidates.  For the F-ratio test, systems that have p-values below 0.02 and that have 10 or more transits are likely to be the most interesting and include (in order of significance) KOIs: 227, 377 (Kepler-9), 806 (Kepler-30), 142, 277 (Kepler-36), 1573, 448, and 918---while still others have undoubtedly interesting signals.

A second analysis for the individual KOIs was to record the ratio of the amplitude of the best fitting sinusoidal model to the uncertainty in that amplitude (the signal-to-noise ratio (SNR) of the sinusoid amplitude fit).  This SNR statistic is given by:
\begin{equation}
SNR = \left( (A^2 + B^2)(\sigma_A^{-2} + \sigma_B^{-2}) \right)^{1/2}
\label{snrequation}
\end{equation}
where $A$ and $B$ are the amplitudes of the best fitting sine and cosine components and $\sigma_A$ and $\sigma_B$ are the uncertainties in those amplitudes (cf., equation (\ref{fitmodel}) below).

\begin{figure}
\includegraphics[width=0.45\textwidth]{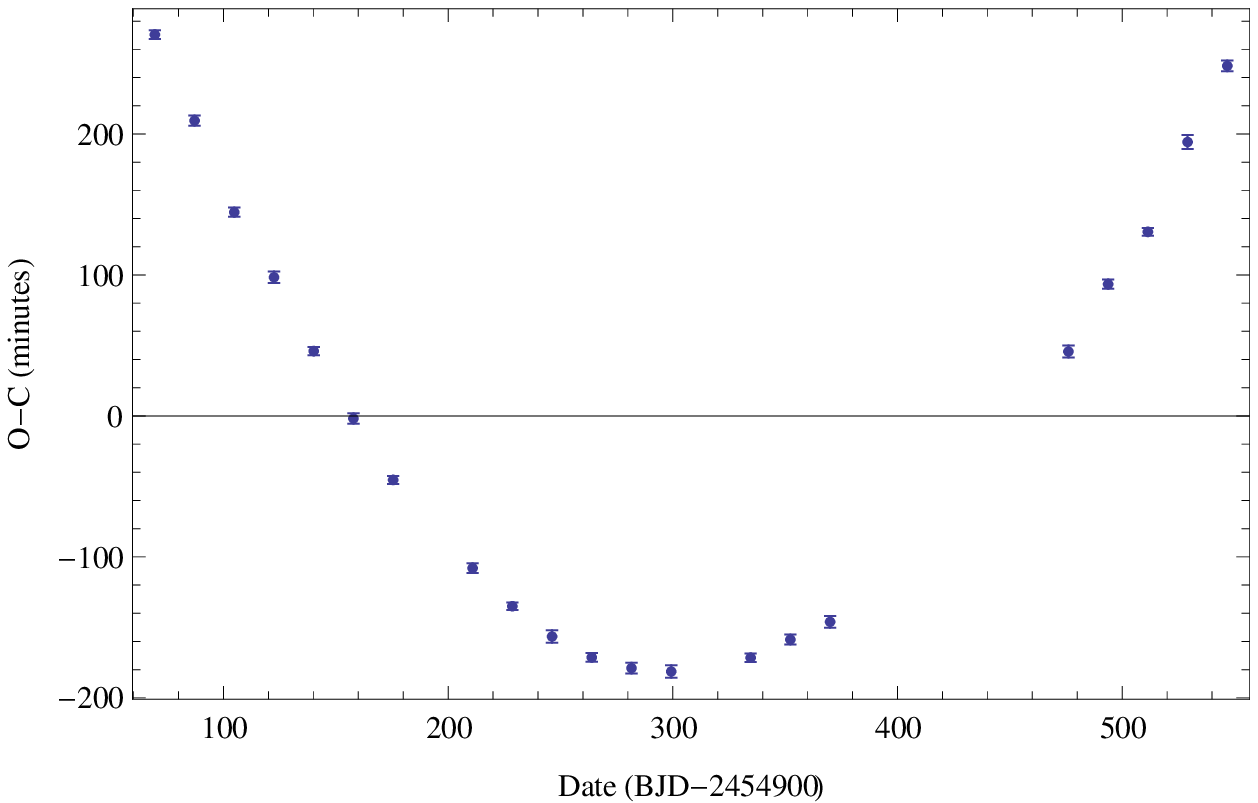}
\includegraphics[width=0.45\textwidth]{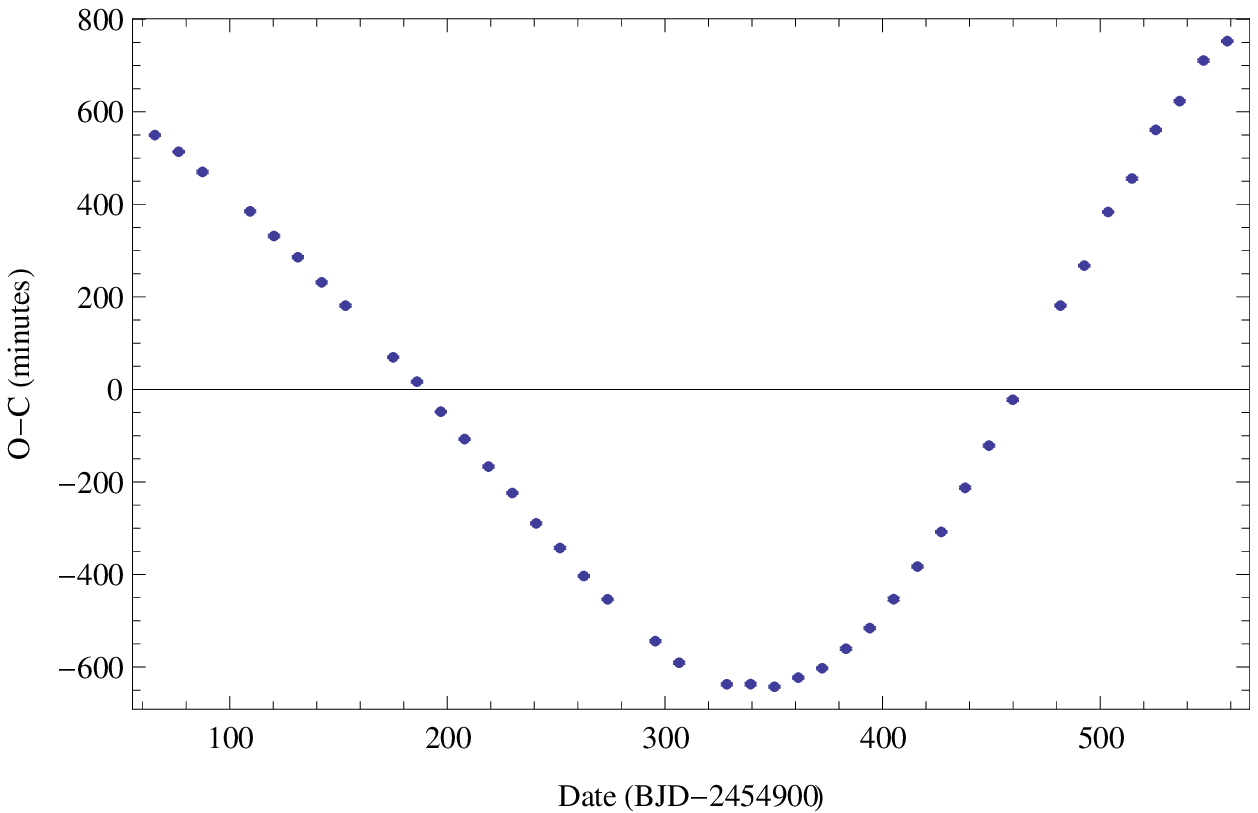}
\includegraphics[width=0.45\textwidth]{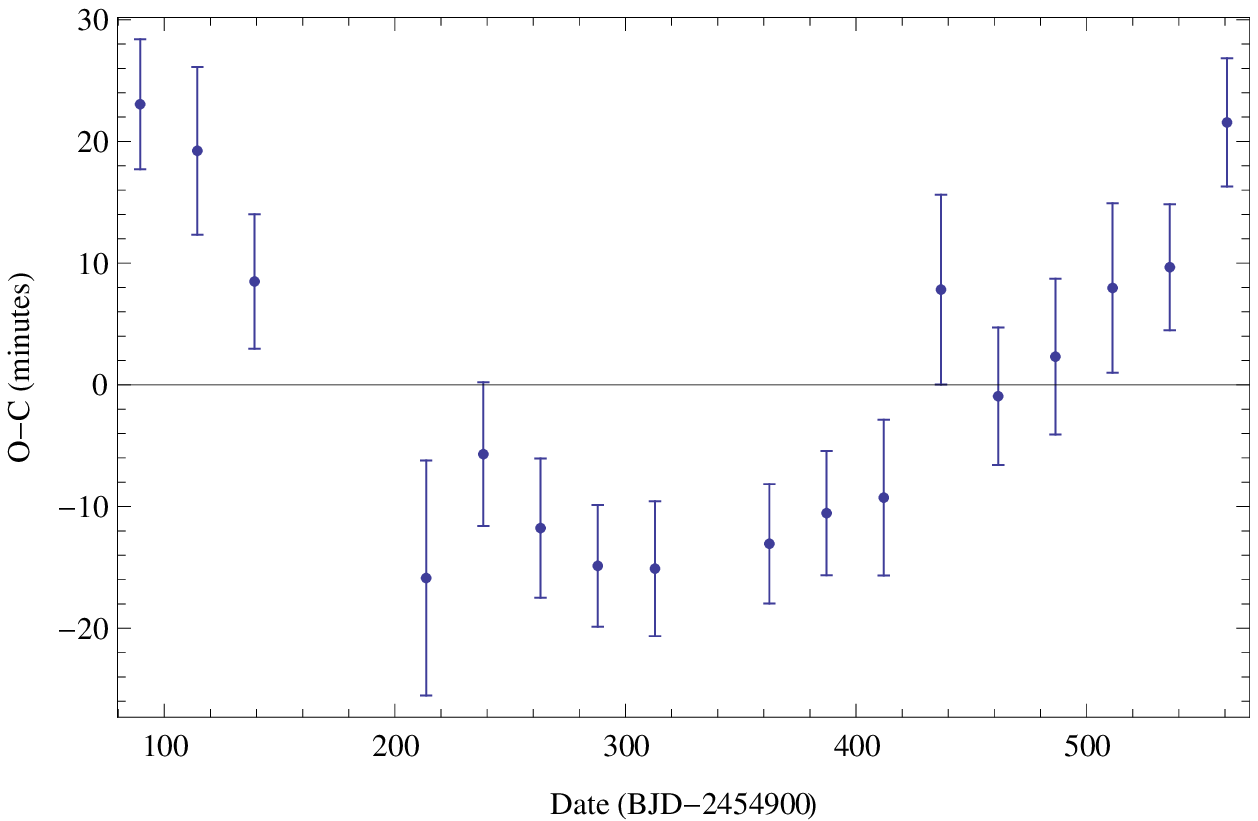}
\caption{Residuals from a linear ephemeris (observed minus calculated) for the three planet candidates with the most significant results from the SNR test and the F-ratio test that have not yet been claimed as planets and that have more than 10 measured transit times.  KOI-227.01 is the top panel, KOI-142.01 is the middle panel, and KOI-1573.01 is the lower panel.\label{examples}}
\end{figure}

Figure \ref{snrplot} shows the results of this test for all KOIs.  As with the F-ratio test, the SNR test shows several systems with sinusoidal amplitudes detected with a large SNR---including many that have not been investigated in depth.  The top candidates from this test with SNR greater than 20 and more than 10 transits include (in order of significance) KOIs: 227 (Kepler-36), 377 (Kepler-9), 142, 1573, 806, 984, 13, 277, 473, and 137 (Kepler-18).  It is noteworthy that not all of the top candidates from the F-ratio test are top candidates for the SNR test and vice versa.  We also point out that in some cases the tests identify some clearly bad candidate systems.  For example, the SNR test identifies KOI-13 with high significance---while its timing residuals are caused by an interesting mix of stellar variability and planet-star dynamics, but not an additional planet \citep{Szabo:2011}.  Figure \ref{examples} shows the TTV signal for the three most significant candidates for both the SNR test and the F-ratio test provided they have more than 10 transit times (KOIs 227, 142, and 1573).  None of these have yet been claimed as confirmed planets.

\begin{figure}
\includegraphics[width=0.45\textwidth]{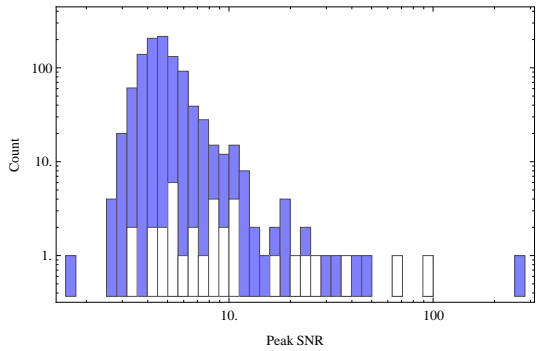}
\caption{Histogram of the SNR values resulting from the SNR test.  Known multibody systems are the light colored bars.  Again, the two most significant systems with this test are KOI-227 and Kepler-9.  However, KOI-142 and KOI-1573 also have very large SNR detections.\label{snrplot}}
\end{figure}

\section{Object Pairs}

For pairs of planets, an additional diagnostic is to look for anticorrelation in the TTV signal.  As planets interact with each other we expect many of their TTV signatures to be anticorrelated due to basic effects from conservation of energy and angular momentum---especially when a system is near MMR.  It is possible to have short-timescale correlated TTV signals in systems farther from resonance or when there is significant precession in one of the objects \citep{Steffen:2012a,Fabrycky:2012,Ford:2012a}.  Using an anticorrelation measurement can have much more power to distinguish real TTV signals from noise as one would expect timing noise to be positively correlated at best and uncorrelated at worst for minimally-interacting planets (some dynamical scenarios can produce positively correlated residuals on short timescales).  Following the procedure outlined in \citep{Steffen:2012a} we calculate \ximax\ for each pair of KOIs that satisfy the selection criteria outlined above for all of the systems with multiple transiting objects.  This is \multisystems\ systems with \multipairs\ total pairs that meet the selection criteria.  The statistic \ximax\ is found from a periodogram generated by fitting the TTV signal to a set of sinusoidal functions with specific periods:
\begin{equation}
f_i = A \sin \left(\frac{2\pi t}{P_i} \right) + B \cos \left(\frac{2\pi t}{P_i} \right) + C
\label{fitmodel}
\end{equation}
where $A$, $B$, and $C$ are model parameters, and $P_i$ is the test timescale\footnote{There is a typo in \citep{Steffen:2012a} where the period of the planet $P$ is in the denominator of the argument of the sine and cosine functions.  This error has been corrected here.}.

The fitted values for $A$ and $B$ and their measured uncertainties, $\sigma_A$ and $\sigma_B$ are then used to calculate \xxii\ for each of the sampled periods using
\begin{equation}
\Xi = -\left(\frac{A_1A_2}{\sigma_{A1}\sigma_{A2}} + \frac{B_1B_2}{\sigma_{B1}\sigma_{B2}}\right)
\label{ximax}
\end{equation}
where the ``1'' and ``2'' subscripts correspond to the two objects.  Finally, the maximum value that \xxii\ has for a given candidate pair, \ximax, is the statistic we choose to determine the significance of any interaction between the two objects.  The square root of \ximax\ is approximately the significance in terms of $\sigma$ for Gaussian deviates, though we note that an accurate determination of the significance is best done via Monte Carlo simulation \citep[see][]{Ford:2012a,Fabrycky:2012,Steffen:2012a}.

\begin{figure}[!ht]
\includegraphics[width=0.45\textwidth]{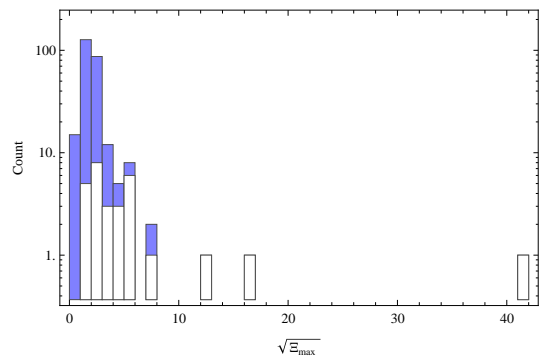}
\caption{Histogram of $\sqrt{\Xi_{\text{max}}}$ for all pairs of candidates.  Pairs from well studied multiobject systems are shown with light-colored bars.\label{ximaxplot}}
\end{figure}

Figure \ref{ximaxplot} shows the results of this analysis for all KOI pairs.  Many of the most significant pairs are in previously identified systems (denoted by the light bars) with Kepler-9 and Kepler-18 being the most significant detections.  Still, several have not been previously identified in the literature including KOI-904.  Other significant pairs may be found by including systems with fewer transits (e.g., Kepler-30) or with transits detected with less significance, however expanding the analysis down to such systems results in many more spurious detections.  The example of KOI-904 shows how correlation statistics have more power to identify potential TTV systems than simply tests on individual KOIs---the KOI-904 system ranks highly with the correlation test, but the individual objects barely make the top 100 for the F-ratio test and are roughly 50th and 100th for the SNR test.  Table \ref{bigoltable} includes a subset of all calculated \ximax\ values---specifically, the \ximax\ value found for a given candidate and its neighbor with the next largest orbital period that satisfies the sample selection criteria.

The companion paper by \citet{Ford:2012b} has similar tests using simple parametric modeling such as quadratic or cubic fits to the TTV signals.  Several of the strongest systems using those methods do not appear strong with the Fourier methods and vice-versa---and the relative ranking is quite different.  This is to be expected as only in limiting cases do the two models have similar functional form.  Moreover, trigonometric functions require a specific relationship between the polynomial powers which generic power-law models do not.  Trigonometric functions have a strong physical motivation (as they are the natural basis to solve the equations of motions of gravitating systems) and they allow for more straightforward correlation tests at specific frequencies as is done in this section.  Nevertheless, the power-law models of \citet{Ford:2012b} can be a faster means of finding interesting systems---especially when there are fewer data to analyze.  The full power of the Fourier methods are best realized as multiple cycles exist in the TTV residuals, as will be increasingly the case as more \kepler\ data come available.

\section{Conclusions}

With the public release of \kepler\ data through Q6, many opportunities are available to study the detailed dynamics of \kepler\ candidate systems.  We have conducted some straightforward statistical tests in an effort to identify systems that may be particularly fruitful.  Some of the most interesting systems have been identified previously.  However, many remain unexplored.  Table \ref{bigoltable} gives the results of each of these tests and can be used to identify systems that are worthy of additional scrutiny.

\kepler 's unique data have enabled unprecedented scientific advances in the study of exoplanets and TTV analyses have played a central role in this endeavor.  With these new data and the identification of additional TTV candidate systems several important opportunities are enabled such as the measurement of planetary masses, dynamical studies of mean-motion resonance, the constraints on mutual inclinations, and the discovery of non-transiting exoplanets.  These can, in turn, inform models of planet formation and dynamical evolution.  Ultimately, TTV analysis is the only means currently available that can measure planet masses of terrestrial planets in the habitable zone of a Sun-like star.  At present, its full potential has not been realized.

\begin{deluxetable}{rrrrrr}
\tablecolumns{6}
\tabletypesize{\normalsize}
\tablewidth{0pc}
\tablecaption{$^a$Period, number of transits, the results of the SNR of the best-fit sinusoidal amplitude, the p-value for the F-ratio test, and \ximax\ for each KOI.\label{bigoltable}}
\tablehead{
\colhead{KOI}  &  \colhead{Period (days)}  &  \colhead{$N_t$}  &  \colhead{SNR}  &  \colhead{p-value}  & \colhead{\ximax$^b$}
}
\startdata
1.01 & 2.471 & 149 & 5.434 & 0.4032 &   \\
2.01 & 2.205 & 208 & 8.227 & 0.3854 &   \\
3.01 & 4.888 & 77 & 6.868 & 0.3426 &   \\
4.01 & 3.849 & 84 & 4.742 & 0.3937 &   \\
5.01 & 4.78 & 87 & 4.3 & 0.3976 &   \\
7.01 & 3.214 & 116 & 4.571 & 0.4045 &   \\
10.01 & 3.522 & 127 & 3.922 & 0.4126 &   \\
12.01 & 17.855 & 24 & 4.152 & 0.3256 &   \\
13.01 & 1.764 & 218 & 28.889 & 0.1449 &   \\
17.01 & 3.235 & 142 & 4.515 & 0.4091 &   \\
18.01 & 3.548 & 127 & 5.887 & 0.3914 &   \\
20.01 & 4.438 & 84 & 5.229 & 0.3829 &   \\
22.01 & 7.891 & 57 & 5.444 & 0.3555 &   \\
41.01 & 12.816 & 35 & 3.725 & 0.3715 & 2.711 \\
41.02 & 6.887 & 62 & 4.424 & 0.3818 & 6.4 \\
41.03 & 35.333 & 12 & 3.479 & 0.2422 &   \\
42.01 & 17.834 & 27 & 6.33 & 0.2423 &   \\
44.01 & 66.468 & 8 & 3.913 & 0.1677 &   \\
46.01 & 3.488 & 132 & 4.4 & 0.4074 &   \\
49.01 & 8.314 & 51 & 5.031 & 0.3599 &   \\
51.01 & 10.431 & 34 & 4.201 & 0.3501 &   \\
63.01 & 9.434 & 48 & 3.335 & 0.3922 &   \\
64.01 & 1.951 & 215 & 5.474 & 0.4106 &   \\
69.01 & 4.727 & 98 & 4.854 & 0.3954 &   \\
70.01 & 10.854 & 40 & 3.731 & 0.3782 & 4.685 \\
70.02 & 3.696 & 120 & 6.514 & 0.3831 & 5.301 \\
70.05 & 19.578 & 23 & 5.163 & 0.2895 &   \\
72.01 & 0.837 & 432 & 5.413 & 0.4211 & 0.676 \\
72.02 & 45.294 & 10 & 4.663 & 0.1601 &   \\
82.01 & 16.146 & 26 & 4.685 & 0.3231 & 2.577 \\
82.02 & 10.312 & 40 & 4.867 & 0.3538 & 3.273 \\
82.03 & 27.453 & 17 & 6.311 & 0.1795 &   \\
84.01 & 9.287 & 47 & 10.965 & 0.197 &   \\
85.01 & 5.86 & 77 & 3.937 & 0.4 & 1.737 \\
85.02 & 2.155 & 185 & 5.572 & 0.4056 & 3.07 \\
85.03 & 8.131 & 56 & 5.517 & 0.3532 &   \\
92.01 & 65.705 & 7 & 8.715 & 0.0202 &   \\
94.01 & 22.343 & 9 & 6.797 & 0.08 &   \\
94.02 & 10.424 & 20 & 7.107 & 0.2132 & 2.945 \\
94.04 & 3.743 & 50 & 4.627 & 0.3626 & 1.151 \\
97.01 & 4.885 & 91 & 4.002 & 0.4034 &   \\
98.01 & 6.79 & 65 & 4.241 & 0.389 &   \\
100.01 & 9.966 & 44 & 5.406 & 0.3331 &   \\
102.01 & 1.735 & 253 & 5.612 & 0.4124 &   \\
103.01 & 14.911 & 29 & 18.727 & 0.0314 &   \\
104.01 & 2.508 & 166 & 5.416 & 0.4048 &   \\
105.01 & 8.981 & 43 & 4.571 & 0.3659 &   \\
107.01 & 7.257 & 63 & 3.379 & 0.4035 &   \\
108.01 & 15.965 & 26 & 3.972 & 0.3443 &   \\
110.01 & 9.941 & 46 & 5.351 & 0.3484 &   \\
111.01 & 11.428 & 39 & 4.12 & 0.3654 & 2.746 \\
111.02 & 23.669 & 19 & 3.017 & 0.3551 & 8.086 \\
111.03 & 51.757 & 6 & 6.253 & 0.0417 &   \\
112.01 & 51.079 & 9 & 7.046 & 0.0608 &   \\
112.02 & 3.709 & 115 & 4.852 & 0.4018 & 1.492 \\
115.01 & 5.412 & 79 & 5.253 & 0.3814 & 5.958 \\
115.02 & 7.126 & 63 & 4.299 & 0.3856 &   \\
116.01 & 13.571 & 33 & 3.985 & 0.3644 & 2.474 \\
116.02 & 43.845 & 11 & 5.062 & 0.1557 &   \\
117.01 & 14.749 & 30 & 4.738 & 0.3597 &   \\
117.02 & 4.901 & 88 & 5.446 & 0.3826 & 5.657 \\
117.03 & 3.18 & 140 & 4.693 & 0.4086 & 5.176 \\
118.01 & 24.993 & 19 & 4.461 & 0.2866 &   \\
119.01 & 49.185 & 10 & 5.725 & 0.1164 &   \\
122.01 & 11.523 & 37 & 4.098 & 0.3627 &   \\
123.01 & 6.482 & 70 & 4.861 & 0.3852 & 2.21 \\
123.02 & 21.222 & 22 & 3.737 & 0.3293 &   \\
124.01 & 12.691 & 36 & 3.749 & 0.3804 & 2.512 \\
124.02 & 31.72 & 15 & 4.745 & 0.2266 &   \\
127.01 & 3.579 & 122 & 6.538 & 0.3807 &   \\
128.01 & 4.943 & 88 & 5.743 & 0.3788 &   \\
131.01 & 5.014 & 91 & 4.535 & 0.3966 &   \\
135.01 & 3.024 & 150 & 4.548 & 0.4099 &   \\
137.01 & 7.642 & 65 & 24.769 & 0.0449 & 262.344 \\
137.02 & 14.859 & 33 & 20.704 & 0.0279 &   \\
137.03 & 3.505 & 120 & 5.058 & 0.3999 & 3.854 \\
138.01 & 48.938 & 10 & 4.115 & 0.2169 &   \\
141.01 & 2.624 & 167 & 4.99 & 0.4091 &   \\
142.01 & 10.951 & 41 & 47.818 & 0.0017 &   \\
144.01 & 4.176 & 88 & 6.022 & 0.3737 &   \\
148.01 & 4.778 & 92 & 4.231 & 0.4008 & 16.017 \\
148.02 & 9.674 & 48 & 8.853 & 0.2421 & 1.711 \\
148.03 & 42.896 & 11 & 4.27 & 0.2197 &   \\
149.01 & 14.557 & 29 & 4.701 & 0.3382 &   \\
150.01 & 8.409 & 54 & 4.807 & 0.3701 & 4.551 \\
150.02 & 28.574 & 11 & 5.433 & 0.1846 &   \\
151.01 & 13.447 & 30 & 4.635 & 0.3238 &   \\
152.01 & 52.091 & 10 & 5.213 & 0.1437 &   \\
152.02 & 27.403 & 17 & 6.745 & 0.181 & 1.704 \\
152.03 & 13.484 & 34 & 3.9 & 0.3582 & 9.329 \\
153.01 & 8.925 & 50 & 5.255 & 0.3552 &   \\
153.02 & 4.754 & 93 & 4.224 & 0.4026 & 3.674 \\
155.01 & 5.661 & 77 & 5.168 & 0.3806 &   \\
156.01 & 8.041 & 52 & 4.974 & 0.361 & 6.293 \\
156.02 & 5.189 & 76 & 5.063 & 0.3811 & 2.211 \\
156.03 & 11.776 & 39 & 4.934 & 0.342 &   \\
157.01 & 13.025 & 33 & 5.389 & 0.3114 & 10.359 \\
157.02 & 22.687 & 19 & 4.448 & 0.2823 & 4.552 \\
157.03 & 31.996 & 13 & 4.14 & 0.2282 & 29.484 \\
157.04 & 46.69 & 9 & 16.842 & 0.0304 &   \\
157.06 & 10.304 & 43 & 5.51 & 0.3354 & 8.485 \\
159.01 & 8.991 & 46 & 4.896 & 0.3504 &   \\
161.01 & 3.106 & 144 & 6.301 & 0.3917 &   \\
162.01 & 14.006 & 31 & 3.628 & 0.3653 &   \\
163.01 & 11.12 & 42 & 4.563 & 0.3575 &   \\
165.01 & 13.222 & 34 & 3.231 & 0.3804 &   \\
166.01 & 12.493 & 34 & 5.716 & 0.3098 &   \\
167.01 & 4.92 & 90 & 4.26 & 0.4008 &   \\
168.01 & 10.742 & 43 & 10.497 & 0.1768 &   \\
171.01 & 5.969 & 73 & 5.215 & 0.3809 & 1.683 \\
171.02 & 13.071 & 34 & 5.413 & 0.3256 &   \\
172.01 & 13.722 & 33 & 4.215 & 0.3529 &   \\
173.01 & 10.061 & 43 & 3.64 & 0.3769 &   \\
174.01 & 56.355 & 8 & 9.311 & 0.0298 &   \\
176.01 & 30.23 & 16 & 4.712 & 0.2442 &   \\
177.01 & 21.061 & 23 & 3.881 & 0.3291 &   \\
179.01 & 20.74 & 21 & 4.058 & 0.3253 &   \\
180.01 & 10.046 & 45 & 6.378 & 0.3118 &   \\
183.01 & 2.684 & 169 & 4.464 & 0.413 &   \\
186.01 & 3.243 & 141 & 5.741 & 0.3953 &   \\
187.01 & 30.883 & 14 & 4.814 & 0.2294 &   \\
188.01 & 3.797 & 118 & 4.239 & 0.4087 &   \\
189.01 & 30.36 & 14 & 4.92 & 0.2213 &   \\
190.01 & 12.265 & 30 & 7.197 & 0.2319 &   \\
191.01 & 15.359 & 25 & 3.702 & 0.347 & -0.005 \\
191.02 & 2.418 & 147 & 4.532 & 0.4098 & 1.488 \\
191.04 & 38.651 & 10 & 6.899 & 0.0955 &   \\
192.01 & 10.291 & 38 & 3.383 & 0.3845 &   \\
193.01 & 37.59 & 13 & 4.31 & 0.2306 &   \\
194.01 & 3.121 & 135 & 5.25 & 0.3998 &   \\
195.01 & 3.218 & 128 & 3.92 & 0.4129 &   \\
196.01 & 1.856 & 243 & 5.441 & 0.4127 &   \\
197.01 & 17.276 & 26 & 6.53 & 0.2479 &   \\
199.01 & 3.269 & 138 & 4.445 & 0.4092 &   \\
200.01 & 7.341 & 57 & 4.367 & 0.382 &   \\
201.01 & 4.225 & 100 & 4.496 & 0.4015 &   \\
202.01 & 1.721 & 254 & 5.322 & 0.4145 &   \\
203.01 & 1.486 & 302 & 4.166 & 0.4223 &   \\
204.01 & 3.247 & 131 & 5.025 & 0.4027 &   \\
205.01 & 11.72 & 39 & 3.682 & 0.3829 &   \\
206.01 & 5.334 & 86 & 4.788 & 0.3903 &   \\
208.01 & 3.004 & 122 & 4.624 & 0.406 &   \\
209.01 & 50.791 & 7 & 7.197 & 0.0166 &   \\
209.02 & 18.796 & 24 & 4.336 & 0.3265 & 0.956 \\
212.01 & 5.696 & 48 & 4.03 & 0.3818 &   \\
214.01 & 3.312 & 115 & 5.248 & 0.3964 &   \\
216.01 & 20.172 & 21 & 4.211 & 0.2808 &   \\
217.01 & 3.905 & 117 & 5.48 & 0.3934 &   \\
219.01 & 8.025 & 56 & 5.05 & 0.3662 &   \\
220.01 & 2.422 & 189 & 5.104 & 0.4095 &   \\
221.01 & 3.413 & 131 & 4.134 & 0.4115 &   \\
222.01 & 6.312 & 66 & 3.888 & 0.3974 & 4.62 \\
222.02 & 12.795 & 30 & 3.918 & 0.3488 &   \\
223.01 & 3.177 & 136 & 4.61 & 0.4061 & 0.57 \\
223.02 & 41.007 & 10 & 3.358 & 0.235 &   \\
225.01 & 0.839 & 335 & 7.091 & 0.4099 &   \\
226.01 & 8.309 & 57 & 4.485 & 0.379 &   \\
227.01 & 17.679 & 21 & 262.562 &   $7.8356\times 10^{-7}$ &   \\
229.01 & 3.573 & 126 & 4.189 & 0.4095 &   \\
232.01 & 12.466 & 37 & 2.861 & 0.3972 & 3.915 \\
232.02 & 5.766 & 77 & 4.91 & 0.3838 & 1.505 \\
232.03 & 21.588 & 20 & 4.676 & 0.2658 & 2.123 \\
232.04 & 37.997 & 11 & 6.816 & 0.1006 & 10.749 \\
232.05 & 56.255 & 9 & 3.24 & 0.2405 &   \\
234.01 & 9.614 & 52 & 4.919 & 0.3664 &   \\
235.01 & 5.633 & 81 & 4.594 & 0.3919 &   \\
237.01 & 8.508 & 51 & 4.508 & 0.3711 &   \\
238.01 & 17.232 & 26 & 3.6 & 0.3572 &   \\
239.01 & 5.641 & 74 & 4.799 & 0.3869 &   \\
240.01 & 4.287 & 106 & 4.495 & 0.4017 &   \\
241.01 & 13.821 & 28 & 6.24 & 0.26 &   \\
242.01 & 7.258 & 62 & 4.568 & 0.3819 &   \\
244.01 & 12.72 & 32 & 4.943 & 0.3163 &   \\
244.02 & 6.239 & 69 & 9.525 & 0.2574 & 25.533 \\
245.01 & 39.792 & 12 & 4.023 & 0.2399 & 1.861 \\
245.02 & 21.302 & 18 & 4.235 & 0.2833 & 6.303 \\
245.04 & 51.198 & 6 & 3.762 & 0.04 &   \\
246.01 & 5.399 & 65 & 4.286 & 0.3877 &   \\
247.01 & 13.815 & 24 & 3.841 & 0.3205 &   \\
248.01 & 7.204 & 62 & 11.535 & 0.2168 & 54.409 \\
248.02 & 10.913 & 37 & 9.913 & 0.1998 &   \\
249.01 & 9.549 & 41 & 3.959 & 0.369 &   \\
250.01 & 12.283 & 34 & 6.503 & 0.285 & 24.613 \\
250.02 & 17.251 & 24 & 5.075 & 0.2774 & 1.428 \\
250.04 & 46.828 & 9 & 3.469 & 0.2354 &   \\
251.01 & 4.164 & 100 & 4.249 & 0.4038 &   \\
252.01 & 17.604 & 24 & 3.791 & 0.3284 &   \\
253.01 & 6.383 & 57 & 4.462 & 0.3797 &   \\
254.01 & 2.455 & 174 & 7.189 & 0.3885 &   \\
255.01 & 27.522 & 17 & 3.309 & 0.3347 &   \\
256.01 & 1.379 & 233 & 6.551 & 0.4045 &   \\
257.01 & 6.883 & 51 & 4.583 & 0.3655 &   \\
258.01 & 4.158 & 88 & 6.239 & 0.371 &   \\
260.01 & 10.496 & 36 & 5.289 & 0.3314 &   \\
261.01 & 16.238 & 27 & 6.192 & 0.2619 &   \\
262.01 & 7.813 & 56 & 4.01 & 0.388 & 6.515 \\
262.02 & 9.376 & 45 & 4.141 & 0.3727 &   \\
263.01 & 20.719 & 18 & 3.107 & 0.3453 &   \\
269.01 & 18.011 & 22 & 6.069 & 0.2336 &   \\
270.01 & 12.583 & 34 & 3.954 & 0.3546 & 3.182 \\
270.02 & 33.673 & 12 & 4.472 & 0.2303 &   \\
271.01 & 48.631 & 8 & 4.211 & 0.1331 &   \\
271.02 & 29.392 & 16 & 4.507 & 0.2536 & 6.753 \\
273.01 & 10.574 & 38 & 6.158 & 0.2923 &   \\
275.01 & 15.792 & 20 & 3.442 & 0.3389 &   \\
276.01 & 41.746 & 11 & 3.787 & 0.2364 &   \\
277.01 & 16.228 & 29 & 27.532 & 0.0079 &   \\
279.01 & 28.455 & 15 & 6.483 & 0.153 &   \\
279.02 & 15.413 & 24 & 4.031 & 0.3267 & 2.809 \\
280.01 & 11.873 & 36 & 3.063 & 0.3887 &   \\
281.01 & 19.557 & 19 & 7.547 & 0.2415 &   \\
282.01 & 27.508 & 14 & 4.884 & 0.2091 &   \\
283.01 & 16.092 & 27 & 4.091 & 0.3361 &   \\
284.01 & 18.01 & 19 & 4.484 & 0.2789 &   \\
284.02 & 6.415 & 49 & 4.411 & 0.3665 & 2.715 \\
284.03 & 6.178 & 54 & 3.897 & 0.3896 & 4.607 \\
285.01 & 13.749 & 33 & 4.761 & 0.3392 &   \\
288.01 & 10.275 & 37 & 5.953 & 0.3091 &   \\
289.01 & 26.629 & 13 & 4.156 & 0.2629 &   \\
291.01 & 31.518 & 15 & 4.542 & 0.2469 &   \\
292.01 & 2.587 & 143 & 5.061 & 0.4013 &   \\
294.01 & 34.436 & 11 & 3.895 & 0.2332 &   \\
295.01 & 5.317 & 68 & 5.766 & 0.3595 &   \\
296.01 & 28.862 & 11 & 3.434 & 0.2541 &   \\
297.01 & 5.652 & 67 & 4.452 & 0.381 &   \\
298.01 & 19.964 & 22 & 3.251 & 0.3562 &   \\
299.01 & 1.542 & 256 & 4.621 & 0.4184 &   \\
301.01 & 6.003 & 73 & 5.231 & 0.3797 &   \\
302.01 & 24.855 & 11 & 3.773 & 0.2585 &   \\
303.01 & 60.929 & 6 & 5.541 & 0.0464 &   \\
304.01 & 8.512 & 48 & 3.838 & 0.382 &   \\
305.01 & 4.604 & 90 & 4.242 & 0.3998 &   \\
306.01 & 24.308 & 15 & 5.159 & 0.2297 &   \\
307.01 & 19.674 & 25 & 3.928 & 0.3339 &   \\
308.01 & 35.595 & 12 & 13.358 & 0.0315 &   \\
312.01 & 11.579 & 37 & 7.49 & 0.2575 & 3.893 \\
312.02 & 16.399 & 27 & 4.914 & 0.2998 &   \\
313.01 & 18.736 & 20 & 4.091 & 0.3135 &   \\
313.02 & 8.436 & 51 & 5.866 & 0.3329 & 8.154 \\
314.01 & 13.781 & 31 & 5.212 & 0.3352 & 4.86 \\
314.02 & 23.09 & 19 & 3.973 & 0.314 &   \\
315.01 & 35.591 & 11 & 3.851 & 0.2505 &   \\
317.01 & 22.208 & 19 & 3.956 & 0.3029 &   \\
318.01 & 38.584 & 11 & 4.927 & 0.1716 &   \\
319.01 & 46.151 & 10 & 9.571 & 0.0327 &   \\
321.01 & 2.426 & 180 & 4.997 & 0.4105 &   \\
323.01 & 5.836 & 71 & 5.593 & 0.3681 &   \\
330.01 & 7.974 & 54 & 3.539 & 0.3936 &   \\
331.01 & 18.684 & 24 & 4.004 & 0.3381 &   \\
332.01 & 5.459 & 81 & 5.742 & 0.3749 &   \\
333.01 & 13.285 & 31 & 4.198 & 0.3337 &   \\
335.01 & 46.567 & 9 & 6.244 & 0.0859 &   \\
337.01 & 19.784 & 24 & 4.252 & 0.3185 &   \\
338.01 & 7.011 & 57 & 4.417 & 0.3776 &   \\
339.01 & 1.98 & 196 & 4.328 & 0.4159 & 1.721 \\
339.02 & 12.834 & 34 & 5.345 & 0.3146 &   \\
340.01 & 23.673 & 16 & 4.856 & 0.2379 &   \\
341.01 & 7.171 & 58 & 3.386 & 0.3992 &   \\
343.01 & 4.762 & 90 & 4.576 & 0.3977 & 4.043 \\
343.02 & 2.024 & 204 & 5.826 & 0.406 & 6.136 \\
343.03 & 41.81 & 10 & 4.837 & 0.1532 &   \\
344.01 & 39.309 & 12 & 4.598 & 0.1963 &   \\
345.01 & 29.886 & 12 & 4.684 & 0.2125 &   \\
346.01 & 12.925 & 34 & 3.491 & 0.3794 &   \\
348.01 & 28.511 & 16 & 5.986 & 0.1743 &   \\
349.01 & 14.387 & 28 & 4.677 & 0.3153 &   \\
350.01 & 12.991 & 28 & 4.013 & 0.344 &   \\
351.03 & 59.74 & 7 & 7.891 & 0.0131 &   \\
352.01 & 27.083 & 17 & 4.073 & 0.2749 &   \\
354.01 & 15.96 & 28 & 4.331 & 0.3358 &   \\
355.01 & 4.903 & 88 & 4.12 & 0.4017 &   \\
356.01 & 1.827 & 237 & 5.891 & 0.409 &   \\
361.01 & 3.248 & 130 & 6.291 & 0.3877 &   \\
366.01 & 75.112 & 6 & 3.314 & 0.1796 &   \\
367.01 & 31.579 & 11 & 5.807 & 0.1375 &   \\
370.01 & 42.882 & 10 & 4.391 & 0.1929 &   \\
370.02 & 22.951 & 18 & 5.78 & 0.2145 & 4.017 \\
377.01 & 19.265 & 24 & 95.814 & 0.0006 & 1755.65 \\
377.02 & 38.882 & 13 & 70.098 & 0.0001 &   \\
379.01 & 6.717 & 66 & 4.18 & 0.3948 &   \\
384.01 & 5.08 & 65 & 4.853 & 0.393 &   \\
385.01 & 13.145 & 34 & 4.592 & 0.3427 &   \\
386.01 & 31.158 & 11 & 6.237 & 0.2416 &   \\
387.01 & 13.9 & 30 & 4.461 & 0.3372 &   \\
388.01 & 6.15 & 65 & 4.482 & 0.3878 &   \\
392.01 & 33.418 & 12 & 7.155 & 0.0949 &   \\
393.01 & 21.416 & 17 & 3.913 & 0.298 &   \\
398.01 & 51.847 & 8 & 4.336 & 0.1388 &   \\
398.02 & 4.18 & 101 & 4.742 & 0.3987 & 3.824 \\
401.01 & 29.199 & 14 & 6.666 & 0.1292 &   \\
403.01 & 21.057 & 19 & 4.165 & 0.2991 &   \\
408.01 & 7.382 & 58 & 4.001 & 0.3927 & 4.004 \\
408.02 & 12.561 & 32 & 3.698 & 0.3719 & 1.627 \\
408.03 & 30.826 & 14 & 5.204 & 0.2104 &   \\
409.01 & 13.249 & 31 & 4.149 & 0.3496 &   \\
410.01 & 7.217 & 58 & 3.797 & 0.3937 &   \\
412.01 & 4.147 & 89 & 4.165 & 0.4007 &   \\
413.01 & 15.229 & 29 & 4.899 & 0.3073 & 2.113 \\
413.02 & 24.675 & 18 & 3.949 & 0.2991 &   \\
416.01 & 18.208 & 20 & 3.883 & 0.3158 & 3.891 \\
416.02 & 88.254 & 6 & 5.653 & 0.0549 &   \\
417.01 & 19.193 & 22 & 4.378 & 0.3138 &   \\
418.01 & 22.418 & 16 & 4.97 & 0.2204 &   \\
419.01 & 20.132 & 21 & 4.436 & 0.293 &   \\
420.01 & 6.01 & 74 & 4.397 & 0.3935 &   \\
421.01 & 4.454 & 101 & 4.893 & 0.397 &   \\
423.01 & 21.087 & 18 & 4.438 & 0.2958 &   \\
425.01 & 5.428 & 71 & 5.676 & 0.376 &   \\
426.01 & 16.301 & 26 & 4.048 & 0.3313 &   \\
427.01 & 24.615 & 16 & 6.981 & 0.2122 & 2.077 \\
427.02 & 42.951 & 12 & 4.15 & 0.2441 &   \\
428.01 & 6.873 & 65 & 3.672 & 0.4 &   \\
429.01 & 8.6 & 55 & 3.271 & 0.3999 &   \\
430.01 & 12.377 & 27 & 2.989 & 0.3737 &   \\
431.01 & 18.87 & 22 & 4.006 & 0.3051 & 1.098 \\
431.02 & 46.903 & 10 & 4.035 & 0.2238 &   \\
432.01 & 5.263 & 83 & 4.011 & 0.4019 &   \\
435.01 & 20.55 & 18 & 4.508 & 0.2888 &   \\
438.01 & 5.931 & 73 & 4.999 & 0.3805 & 2.065 \\
438.02 & 52.663 & 8 & 4.934 & 0.099 &   \\
439.01 & 1.902 & 229 & 4.831 & 0.4156 &   \\
440.01 & 15.907 & 28 & 4.856 & 0.3106 &   \\
440.02 & 4.973 & 85 & 5.297 & 0.3843 & 3.807 \\
442.01 & 13.54 & 33 & 4.173 & 0.356 &   \\
443.01 & 16.218 & 27 & 5.971 & 0.2691 &   \\
444.01 & 11.723 & 33 & 4.306 & 0.3485 &   \\
446.01 & 16.709 & 22 & 4.742 & 0.298 & 6.528 \\
446.02 & 28.551 & 12 & 3.864 & 0.2496 &   \\
448.01 & 10.14 & 43 & 5.773 & 0.3207 & 11.975 \\
448.02 & 43.611 & 10 & 17.921 & 0.0121 &   \\
452.01 & 3.706 & 121 & 3.573 & 0.4145 &   \\
454.01 & 29.007 & 11 & 5.388 & 0.1651 &   \\
456.01 & 13.7 & 30 & 19.024 & 0.2226 &   \\
457.01 & 4.921 & 87 & 4.293 & 0.3977 & 3.419 \\
457.02 & 7.064 & 65 & 7.217 & 0.3198 &   \\
458.01 & 53.718 & 6 & 11.63 & 0.0059 &   \\
459.01 & 19.446 & 19 & 6.004 & 0.242 &   \\
460.01 & 17.587 & 23 & 5.958 & 0.3361 &   \\
463.01 & 18.478 & 22 & 5.395 & 0.2701 &   \\
464.01 & 58.362 & 8 & 7.338 & 0.0408 &   \\
464.02 & 5.35 & 79 & 4.424 & 0.3941 & 1.95 \\
466.01 & 9.391 & 50 & 5.029 & 0.3563 &   \\
467.01 & 18.009 & 23 & 3.932 & 0.3355 &   \\
468.01 & 22.184 & 19 & 4.69 & 0.3059 &   \\
469.01 & 10.329 & 41 & 4.424 & 0.3654 &   \\
470.01 & 3.751 & 113 & 4.89 & 0.3995 &   \\
471.01 & 21.348 & 21 & 4.183 & 0.3139 &   \\
472.01 & 4.244 & 102 & 4.849 & 0.3947 &   \\
473.01 & 12.706 & 35 & 24.84 & 0.1392 &   \\
474.01 & 10.946 & 39 & 3.231 & 0.3882 & 3.203 \\
474.02 & 28.986 & 14 & 4.671 & 0.2434 &   \\
475.01 & 8.181 & 54 & 4.33 & 0.38 & 3.86 \\
475.02 & 15.313 & 29 & 3.566 & 0.3694 &   \\
476.01 & 18.428 & 24 & 3.861 & 0.2986 &   \\
477.01 & 16.543 & 25 & 6.145 & 0.2378 &   \\
478.01 & 11.023 & 39 & 3.612 & 0.3826 &   \\
479.01 & 34.189 & 12 & 5.052 & 0.1697 &   \\
480.01 & 4.302 & 108 & 4.231 & 0.4068 &   \\
481.01 & 7.65 & 58 & 4.468 & 0.3805 & 2.665 \\
481.03 & 34.261 & 12 & 3.5 & 0.2634 &   \\
483.01 & 4.799 & 72 & 3.612 & 0.4028 &   \\
484.01 & 17.205 & 26 & 5.619 & 0.2682 &   \\
486.01 & 22.183 & 20 & 5.768 & 0.2384 &   \\
487.01 & 7.659 & 59 & 4.826 & 0.3775 &   \\
488.01 & 9.379 & 46 & 4.699 & 0.3572 &   \\
490.03 & 7.406 & 59 & 3.967 & 0.3926 &   \\
492.01 & 29.911 & 12 & 4.989 & 0.1917 &   \\
494.01 & 25.697 & 15 & 7.492 & 0.2474 &   \\
496.01 & 1.617 & 253 & 5.056 & 0.4159 &   \\
497.01 & 13.193 & 27 & 4.634 & 0.3185 &   \\
499.01 & 9.668 & 36 & 4.515 & 0.3523 &   \\
500.01 & 7.053 & 49 & 7.014 & 0.3013 & 29.521 \\
500.02 & 9.522 & 38 & 8.892 & 0.2127 &   \\
501.01 & 24.795 & 17 & 3.688 & 0.3152 &   \\
503.01 & 8.222 & 45 & 3.788 & 0.3786 &   \\
504.01 & 40.606 & 6 & 7.74 & 0.0291 &   \\
505.01 & 13.767 & 27 & 6.223 & 0.2491 &   \\
506.01 & 1.583 & 269 & 5.683 & 0.4135 &   \\
507.01 & 18.492 & 21 & 4.219 & 0.304 &   \\
508.01 & 7.931 & 43 & 3.952 & 0.3797 & 2.577 \\
508.02 & 16.666 & 19 & 4.28 & 0.3044 &   \\
509.01 & 4.167 & 103 & 4.348 & 0.4037 & 4.24 \\
509.02 & 11.464 & 38 & 3.683 & 0.3663 &   \\
510.01 & 2.94 & 120 & 4.744 & 0.4006 & 2.448 \\
510.02 & 6.389 & 49 & 3.123 & 0.4005 & 4.536 \\
510.03 & 14.627 & 22 & 4.995 & 0.2817 &   \\
511.01 & 8.006 & 54 & 4.429 & 0.3788 &   \\
512.01 & 6.51 & 66 & 4.557 & 0.3815 &   \\
513.01 & 35.181 & 14 & 3.025 & 0.3407 &   \\
517.01 & 2.752 & 150 & 4.759 & 0.4077 &   \\
518.01 & 13.982 & 27 & 3.56 & 0.3542 & 4.666 \\
518.02 & 44 & 10 & 3.508 & 0.2238 &   \\
519.01 & 11.903 & 35 & 4.139 & 0.3622 & 3.955 \\
519.02 & 34.036 & 14 & 4.292 & 0.2213 &   \\
520.01 & 12.76 & 31 & 3.662 & 0.3653 & 5.972 \\
520.03 & 25.751 & 18 & 4.168 & 0.2913 &   \\
521.01 & 10.161 & 44 & 3.257 & 0.3942 &   \\
522.01 & 12.83 & 33 & 5.912 & 0.2955 &   \\
523.01 & 49.413 & 9 & 4.323 & 0.1843 &   \\
523.02 & 36.854 & 12 & 4.954 & 0.1877 & 4.757 \\
524.01 & 4.592 & 96 & 18.435 & 0.1451 &   \\
525.01 & 11.532 & 38 & 4.748 & 0.3421 &   \\
526.01 & 2.105 & 201 & 4.852 & 0.4127 &   \\
528.01 & 9.577 & 48 & 6.222 & 0.3253 & 1.571 \\
528.02 & 96.672 & 6 & 10.154 & 0.0042 &   \\
528.03 & 20.553 & 22 & 4.287 & 0.3056 & 0.388 \\
530.01 & 10.94 & 37 & 4.147 & 0.3606 &   \\
531.01 & 3.687 & 100 & 3.552 & 0.4115 &   \\
532.01 & 4.222 & 106 & 4.323 & 0.4028 &   \\
533.01 & 16.55 & 27 & 5.241 & 0.2911 &   \\
534.01 & 6.4 & 64 & 3.759 & 0.3966 &   \\
535.01 & 5.853 & 75 & 4.555 & 0.3897 &   \\
537.01 & 2.82 & 135 & 5.046 & 0.4036 &   \\
538.01 & 21.217 & 19 & 3.645 & 0.3286 &   \\
541.01 & 13.646 & 32 & 4.07 & 0.3467 &   \\
542.01 & 41.886 & 11 & 3.768 & 0.2443 &   \\
543.01 & 4.302 & 98 & 3.485 & 0.4105 &   \\
546.01 & 20.684 & 19 & 3.687 & 0.3212 &   \\
547.01 & 25.303 & 19 & 4.359 & 0.2881 &   \\
548.01 & 21.3 & 19 & 4.12 & 0.3098 &   \\
550.01 & 13.024 & 25 & 3.267 & 0.3627 &   \\
551.01 & 11.637 & 32 & 5.204 & 0.3125 &   \\
552.01 & 3.055 & 130 & 5.368 & 0.399 &   \\
554.01 & 3.658 & 121 & 5.126 & 0.3989 &   \\
557.01 & 15.656 & 24 & 3.278 & 0.3649 &   \\
558.01 & 9.178 & 43 & 5.444 & 0.3391 &   \\
560.01 & 23.675 & 17 & 4.34 & 0.2845 &   \\
561.01 & 5.379 & 81 & 4.639 & 0.3892 &   \\
563.01 & 15.284 & 30 & 7.088 & 0.2294 &   \\
564.01 & 21.055 & 19 & 4.625 & 0.2782 &   \\
566.01 & 25.855 & 9 & 5.001 & 0.1607 &   \\
567.01 & 10.688 & 37 & 4.802 & 0.3383 & 5.153 \\
567.02 & 20.303 & 23 & 5.34 & 0.2817 & 1.285 \\
567.03 & 29.023 & 14 & 3.715 & 0.2881 &   \\
569.01 & 20.729 & 20 & 7.87 & 0.1586 &   \\
571.01 & 7.267 & 56 & 6.161 & 0.3433 & 3.026 \\
571.02 & 13.343 & 32 & 3.1 & 0.3594 & 2.437 \\
571.04 & 22.408 & 20 & 3.821 & 0.2801 &   \\
572.01 & 10.64 & 41 & 4.136 & 0.367 &   \\
573.01 & 5.997 & 75 & 4.182 & 0.3938 &   \\
574.01 & 20.135 & 23 & 3.662 & 0.3491 &   \\
575.01 & 24.316 & 16 & 4.787 & 0.2681 &   \\
578.01 & 6.412 & 69 & 3.667 & 0.4022 &   \\
580.01 & 6.521 & 66 & 4.759 & 0.3826 &   \\
581.01 & 6.997 & 60 & 5.155 & 0.3668 &   \\
582.01 & 5.945 & 73 & 4.794 & 0.3872 & 4.649 \\
582.02 & 17.739 & 23 & 4.776 & 0.2972 &   \\
584.01 & 9.927 & 43 & 4.355 & 0.3686 & 2.981 \\
584.02 & 21.223 & 20 & 6.116 & 0.2944 &   \\
585.01 & 3.722 & 114 & 6.379 & 0.3819 &   \\
586.01 & 15.78 & 22 & 3.906 & 0.3449 &   \\
587.01 & 14.035 & 32 & 3.776 & 0.3575 &   \\
588.01 & 10.356 & 41 & 3.872 & 0.3731 &   \\
590.01 & 11.389 & 35 & 3.781 & 0.3732 & 6.259 \\
590.02 & 50.697 & 8 & 5.947 & 0.084 &   \\
592.01 & 39.749 & 12 & 8.804 & 0.0602 &   \\
593.01 & 9.997 & 43 & 3.856 & 0.381 &   \\
597.01 & 17.308 & 22 & 3.188 & 0.3659 &   \\
598.01 & 8.308 & 56 & 4.655 & 0.3715 &   \\
599.01 & 6.454 & 67 & 5.399 & 0.3697 &   \\
600.01 & 3.596 & 106 & 4.409 & 0.4045 &   \\
601.01 & 5.404 & 76 & 3.934 & 0.3979 & 8.225 \\
601.02 & 11.679 & 37 & 6.289 & 0.3089 &   \\
602.01 & 12.914 & 32 & 3.884 & 0.3596 &   \\
605.01 & 2.628 & 163 & 5.092 & 0.4063 &   \\
607.01 & 5.894 & 44 & 6.674 & 0.34 &   \\
609.01 & 4.397 & 91 & 4.987 & 0.3913 &   \\
610.01 & 14.283 & 22 & 4.16 & 0.3237 &   \\
611.01 & 3.252 & 126 & 4.652 & 0.4055 &   \\
612.01 & 20.74 & 16 & 4.428 & 0.2775 & 2.592 \\
612.02 & 47.426 & 7 & 1.624 & 0.0523 &   \\
614.01 & 12.875 & 32 & 5.813 & 0.3008 &   \\
617.01 & 37.865 & 10 & 6.04 & 0.1363 &   \\
618.01 & 9.071 & 48 & 4.386 & 0.3701 &   \\
620.01 & 45.155 & 9 & 5.447 & 0.1377 &   \\
623.01 & 10.35 & 42 & 4.591 & 0.3576 & 4.466 \\
623.02 & 15.677 & 25 & 4.671 & 0.3118 &   \\
624.01 & 17.79 & 23 & 3.998 & 0.3379 & 3.224 \\
624.02 & 49.567 & 9 & 7.379 & 0.0604 &   \\
625.01 & 38.139 & 10 & 3.473 & 0.2629 &   \\
626.01 & 14.587 & 25 & 4.717 & 0.3113 &   \\
627.01 & 7.752 & 57 & 3.821 & 0.3953 &   \\
628.01 & 14.486 & 32 & 4.516 & 0.3373 &   \\
629.01 & 40.699 & 10 & 3.124 & 0.2853 &   \\
632.01 & 7.239 & 57 & 4.602 & 0.3757 &   \\
635.01 & 16.72 & 20 & 4.117 & 0.3103 &   \\
638.01 & 23.636 & 15 & 2.963 & 0.266 &   \\
639.01 & 17.98 & 20 & 5.272 & 0.273 &   \\
640.01 & 30.997 & 13 & 4.524 & 0.26 &   \\
641.01 & 14.852 & 26 & 4.526 & 0.3343 &   \\
644.01 & 45.978 & 7 & 6.787 & 0.0512 &   \\
645.02 & 23.784 & 16 & 3.733 & 0.3109 &   \\
647.01 & 5.169 & 87 & 3.856 & 0.4058 &   \\
649.01 & 23.45 & 19 & 4.677 & 0.2854 &   \\
650.01 & 11.955 & 38 & 4.525 & 0.3575 &   \\
652.01 & 16.081 & 26 & 5.036 & 0.2958 &   \\
654.01 & 8.594 & 41 & 4.981 & 0.3482 &   \\
655.01 & 25.672 & 17 & 3.199 & 0.3251 &   \\
657.01 & 4.069 & 83 & 6.761 & 0.3554 & 5.688 \\
657.02 & 16.283 & 22 & 4.433 & 0.3112 &   \\
658.01 & 3.163 & 133 & 5.434 & 0.3987 & 5.205 \\
658.02 & 5.371 & 77 & 4.849 & 0.3834 &   \\
659.01 & 23.206 & 19 & 4.821 & 0.2775 &   \\
660.01 & 6.08 & 60 & 4.73 & 0.3744 &   \\
661.01 & 14.401 & 25 & 4.018 & 0.3434 &   \\
662.01 & 10.214 & 43 & 4.588 & 0.361 &   \\
663.01 & 2.756 & 118 & 5.008 & 0.4007 & 2.645 \\
663.02 & 20.307 & 18 & 4.672 & 0.2592 &   \\
664.01 & 13.137 & 34 & 4.82 & 0.3322 &   \\
665.01 & 5.868 & 74 & 4.62 & 0.3887 &   \\
666.01 & 22.248 & 18 & 4.823 & 0.2959 &   \\
667.01 & 4.305 & 97 & 4.748 & 0.3975 &   \\
670.01 & 9.489 & 45 & 4.8 & 0.3518 &   \\
672.01 & 16.088 & 26 & 4.137 & 0.3509 & 0.858 \\
672.02 & 41.749 & 10 & 4.832 & 0.152 &   \\
673.01 & 4.417 & 97 & 3.921 & 0.4065 &   \\
674.01 & 16.339 & 15 & 3.466 & 0.3063 &   \\
676.01 & 7.973 & 55 & 6.682 & 0.3159 &   \\
676.02 & 2.453 & 173 & 3.766 & 0.418 & 2.099 \\
678.01 & 6.04 & 68 & 4.168 & 0.3944 &   \\
678.02 & 4.139 & 93 & 4.825 & 0.3931 & 3.463 \\
679.01 & 31.806 & 12 & 4.685 & 0.2216 &   \\
680.01 & 8.6 & 48 & 4.365 & 0.3686 &   \\
684.01 & 4.035 & 104 & 3.245 & 0.4149 &   \\
685.01 & 3.174 & 134 & 4.728 & 0.4058 &   \\
686.01 & 52.514 & 8 & 5.964 & 0.0952 &   \\
688.01 & 3.276 & 135 & 3.491 & 0.4179 &   \\
689.01 & 15.874 & 15 & 4.043 & 0.1792 &   \\
691.01 & 29.666 & 14 & 4.299 & 0.2464 &   \\
693.01 & 28.779 & 13 & 5.928 & 0.1597 &   \\
693.02 & 15.66 & 27 & 4.739 & 0.3143 & 5.857 \\
694.01 & 17.421 & 23 & 3.234 & 0.3661 &   \\
695.01 & 29.908 & 13 & 4.641 & 0.2468 &   \\
697.01 & 3.032 & 81 & 5.917 & 0.3721 &   \\
698.01 & 12.719 & 32 & 5.434 & 0.3132 &   \\
700.01 & 30.864 & 15 & 3.305 & 0.3225 &   \\
700.02 & 9.361 & 45 & 7.036 & 0.2967 & 2.241 \\
700.03 & 14.667 & 28 & 2.939 & 0.3852 & 3.305 \\
701.01 & 18.164 & 22 & 3.88 & 0.3391 &   \\
701.02 & 5.715 & 75 & 4.874 & 0.3839 & 2.848 \\
704.01 & 18.396 & 20 & 11.217 & 0.3273 &   \\
707.01 & 21.776 & 19 & 4.459 & 0.2886 & 8.616 \\
707.02 & 41.027 & 12 & 4.387 & 0.2234 &   \\
707.03 & 31.785 & 14 & 3.977 & 0.274 & 1.423 \\
707.04 & 13.176 & 34 & 4.006 & 0.3676 & 2.471 \\
708.01 & 17.406 & 23 & 3.263 & 0.3535 &   \\
708.02 & 7.693 & 54 & 4.78 & 0.3691 & 2.776 \\
709.01 & 21.384 & 22 & 4.89 & 0.3088 &   \\
710.01 & 5.375 & 77 & 4.213 & 0.3947 &   \\
711.01 & 44.699 & 9 & 4.135 & 0.2101 &   \\
714.01 & 4.182 & 108 & 3.974 & 0.4073 &   \\
716.01 & 26.893 & 16 & 3.502 & 0.3174 &   \\
717.01 & 14.707 & 29 & 5.545 & 0.2927 &   \\
718.01 & 4.585 & 99 & 4.244 & 0.4029 & 4.306 \\
718.02 & 22.715 & 18 & 4.185 & 0.3086 & 3.011 \\
718.03 & 47.906 & 10 & 5.317 & 0.1197 &   \\
719.01 & 9.034 & 48 & 4.405 & 0.3737 &   \\
720.01 & 5.691 & 68 & 3.836 & 0.3982 & 7.732 \\
720.02 & 10.042 & 43 & 4.834 & 0.3526 & 3.849 \\
720.03 & 18.37 & 23 & 5.289 & 0.2723 &   \\
720.04 & 2.796 & 147 & 5.092 & 0.4022 & 5.038 \\
721.01 & 13.724 & 31 & 5.298 & 0.3142 &   \\
722.01 & 46.408 & 10 & 3.215 & 0.2672 &   \\
723.01 & 3.937 & 102 & 4.841 & 0.398 & 3.294 \\
723.02 & 28.082 & 13 & 7.868 & 0.0853 &   \\
723.03 & 10.089 & 39 & 4.601 & 0.3516 & 4.523 \\
725.01 & 7.305 & 61 & 5.843 & 0.3493 &   \\
728.01 & 7.189 & 59 & 5.765 & 0.3537 &   \\
730.01 & 14.788 & 26 & 3.981 & 0.3432 & 5.444 \\
730.03 & 19.722 & 20 & 4.098 & 0.3121 &   \\
732.01 & 1.26 & 342 & 4.373 & 0.4225 &   \\
733.01 & 5.925 & 72 & 4.869 & 0.3831 & 1.915 \\
733.02 & 11.349 & 37 & 4.131 & 0.3634 &   \\
734.01 & 24.543 & 16 & 5.682 & 0.1727 &   \\
735.01 & 22.341 & 19 & 5.57 & 0.2977 &   \\
736.01 & 18.794 & 22 & 7.622 & 0.3124 &   \\
737.01 & 14.498 & 29 & 10.639 & 0.3277 &   \\
738.01 & 10.338 & 45 & 10.787 & 0.2776 & 26.102 \\
738.02 & 13.291 & 32 & 4.833 & 0.3388 &   \\
740.01 & 17.672 & 24 & 3.938 & 0.3208 &   \\
741.01 & 23.355 & 17 & 4.782 & 0.2526 &   \\
743.01 & 19.404 & 22 & 3.824 & 0.3319 &   \\
745.01 & 16.47 & 26 & 4.2 & 0.3382 &   \\
746.01 & 9.274 & 42 & 3.027 & 0.3965 &   \\
747.01 & 6.029 & 69 & 4.468 & 0.3919 &   \\
749.01 & 5.35 & 80 & 3.793 & 0.4034 &   \\
750.01 & 21.677 & 20 & 6.737 & 0.1621 &   \\
752.01 & 9.488 & 46 & 3.307 & 0.3935 & 2.181 \\
752.02 & 54.415 & 9 & 4.836 & 0.1284 &   \\
753.01 & 19.899 & 22 & 5.215 & 0.2835 &   \\
756.01 & 11.094 & 37 & 3.721 & 0.3722 &   \\
757.01 & 16.069 & 24 & 4.591 & 0.3024 & 4.432 \\
757.02 & 41.192 & 10 & 4.162 & 0.1793 &   \\
757.03 & 6.253 & 59 & 5.513 & 0.3592 & 6.709 \\
758.01 & 16.013 & 22 & 4.486 & 0.3069 &   \\
759.01 & 32.624 & 11 & 4.891 & 0.1692 &   \\
760.01 & 4.959 & 86 & 4.004 & 0.402 &   \\
763.01 & 19.651 & 18 & 5.953 & 0.2147 &   \\
764.01 & 41.439 & 11 & 9.072 & 0.048 &   \\
765.01 & 8.354 & 46 & 4.542 & 0.3358 &   \\
766.01 & 4.126 & 99 & 4.437 & 0.4022 &   \\
767.01 & 2.817 & 156 & 4.211 & 0.4147 &   \\
769.01 & 4.281 & 100 & 4.068 & 0.406 &   \\
773.01 & 38.378 & 8 & 4.075 & 0.1744 &   \\
774.01 & 7.443 & 62 & 6.094 & 0.3483 &   \\
775.01 & 16.385 & 23 & 5.291 & 0.2592 & 10.104 \\
775.02 & 7.877 & 44 & 7.56 & 0.2838 & 12.825 \\
775.03 & 36.446 & 11 & 7.836 & 0.1461 &   \\
776.01 & 3.729 & 122 & 5.037 & 0.3992 &   \\
777.01 & 40.42 & 11 & 6.548 & 0.1032 &   \\
779.01 & 10.406 & 32 & 4.139 & 0.3547 &   \\
780.01 & 2.337 & 174 & 6.287 & 0.399 &   \\
781.01 & 11.598 & 40 & 5.07 & 0.345 &   \\
782.01 & 6.575 & 63 & 6.056 & 0.3506 &   \\
783.01 & 7.275 & 62 & 4.697 & 0.3785 &   \\
784.01 & 19.271 & 20 & 16.52 & 0.059 &   \\
785.01 & 12.393 & 32 & 5.051 & 0.3281 &   \\
787.01 & 4.431 & 97 & 4.83 & 0.3946 & 4.005 \\
787.02 & 11.379 & 39 & 4.036 & 0.3676 &   \\
788.01 & 26.394 & 17 & 2.686 & 0.3523 &   \\
790.01 & 8.472 & 45 & 3.439 & 0.3936 &   \\
791.01 & 12.612 & 35 & 2.944 & 0.3911 &   \\
795.01 & 6.77 & 63 & 5.101 & 0.3724 &   \\
797.01 & 10.182 & 42 & 3.81 & 0.3802 &   \\
799.01 & 1.627 & 222 & 5.982 & 0.4061 &   \\
800.01 & 2.711 & 159 & 4.314 & 0.4118 & 2.096 \\
800.02 & 7.212 & 53 & 3.466 & 0.3961 &   \\
801.01 & 1.626 & 219 & 4.885 & 0.4151 &   \\
802.01 & 19.62 & 17 & 4.214 & 0.2971 &   \\
804.01 & 9.029 & 46 & 5.004 & 0.3575 &   \\
805.01 & 10.328 & 32 & 4.116 & 0.3564 &   \\
806.02 & 60.325 & 6 & 8.23 & 0.0095 &   \\
806.03 & 29.287 & 15 & 39.307 & 0.0008 & 153.317 \\
809.01 & 1.595 & 201 & 5.862 & 0.4053 &   \\
810.01 & 4.783 & 92 & 4.658 & 0.3967 &   \\
811.01 & 20.506 & 22 & 4.79 & 0.2988 &   \\
812.01 & 3.34 & 127 & 4.818 & 0.4041 & 1.679 \\
812.02 & 20.061 & 23 & 3.888 & 0.3447 & 3.818 \\
812.03 & 46.183 & 8 & 6.12 & 0.0684 &   \\
813.01 & 3.896 & 93 & 6.302 & 0.3656 &   \\
814.01 & 22.366 & 16 & 2.781 & 0.2695 &   \\
815.01 & 34.844 & 14 & 7.125 & 0.1275 &   \\
816.01 & 7.748 & 38 & 5.887 & 0.3051 &   \\
817.01 & 23.968 & 13 & 9.828 & 0.0572 &   \\
818.01 & 8.114 & 49 & 4.207 & 0.3808 &   \\
821.01 & 21.813 & 13 & 4.736 & 0.2155 &   \\
822.01 & 7.919 & 45 & 4.71 & 0.3646 &   \\
823.01 & 1.028 & 329 & 10.252 & 0.3884 &   \\
824.01 & 15.376 & 6 & 10.833 & 0.0046 &   \\
825.01 & 8.104 & 38 & 5.804 & 0.3095 &   \\
826.01 & 6.366 & 73 & 3.667 & 0.4009 &   \\
827.01 & 5.976 & 62 & 3.741 & 0.3923 &   \\
829.01 & 18.648 & 23 & 5.7 & 0.2514 & 2.835 \\
829.03 & 38.561 & 12 & 3.293 & 0.2924 &   \\
830.01 & 3.526 & 126 & 4.889 & 0.4016 &   \\
833.01 & 3.951 & 88 & 4.135 & 0.404 &   \\
834.01 & 23.654 & 19 & 6.106 & 0.2211 &   \\
834.02 & 13.234 & 31 & 4.874 & 0.3223 & 1.509 \\
835.01 & 11.763 & 29 & 5.907 & 0.2661 & 8.451 \\
835.02 & 56.229 & 8 & 4.448 & 0.1496 &   \\
837.01 & 7.954 & 48 & 3.967 & 0.3823 &   \\
838.01 & 4.859 & 93 & 4.266 & 0.3984 &   \\
840.01 & 3.04 & 144 & 5.697 & 0.3979 &   \\
841.01 & 15.335 & 29 & 8.641 & 0.1753 & 19.262 \\
841.02 & 31.33 & 11 & 5.067 & 0.1763 &   \\
842.01 & 12.718 & 30 & 3.973 & 0.3607 & 0.696 \\
842.02 & 36.065 & 12 & 4.813 & 0.2082 &   \\
843.01 & 4.19 & 107 & 5.471 & 0.3873 &   \\
844.01 & 3.71 & 93 & 4.413 & 0.4 &   \\
845.01 & 16.33 & 22 & 4.345 & 0.3075 &   \\
846.01 & 27.808 & 16 & 6.215 & 0.1665 &   \\
847.01 & 80.871 & 6 & 4.885 & 0.0668 &   \\
849.01 & 10.355 & 45 & 5.258 & 0.3417 &   \\
850.01 & 10.526 & 45 & 3.61 & 0.3873 &   \\
851.01 & 4.584 & 98 & 4.081 & 0.4055 &   \\
853.01 & 8.204 & 53 & 3.641 & 0.392 & 10.23 \\
853.02 & 14.497 & 30 & 7.05 & 0.2808 &   \\
854.01 & 56.056 & 8 & 6.722 & 0.0487 &   \\
855.01 & 41.408 & 10 & 5.953 & 0.1243 &   \\
856.01 & 39.749 & 11 & 3.607 & 0.2669 &   \\
857.01 & 5.715 & 61 & 4.209 & 0.388 &   \\
858.01 & 13.61 & 28 & 3.621 & 0.354 &   \\
864.01 & 4.312 & 91 & 4.269 & 0.4029 & 5.52 \\
864.02 & 20.05 & 22 & 4.167 & 0.3154 &   \\
867.01 & 16.085 & 28 & 4.992 & 0.3026 &   \\
869.01 & 7.49 & 59 & 4.529 & 0.3823 & 3.201 \\
869.02 & 36.285 & 12 & 3.181 & 0.223 &   \\
869.03 & 17.461 & 24 & 5.69 & 0.2667 & 0.978 \\
870.01 & 5.912 & 72 & 10.823 & 0.2513 & 29.827 \\
870.02 & 8.986 & 49 & 5.827 & 0.3421 &   \\
871.01 & 12.941 & 30 & 4.545 & 0.3397 &   \\
872.01 & 33.603 & 14 & 12.047 & 0.0313 &   \\
874.01 & 4.602 & 93 & 4.701 & 0.392 &   \\
875.01 & 4.221 & 102 & 3.977 & 0.4085 &   \\
876.01 & 6.998 & 56 & 3.714 & 0.3916 &   \\
877.01 & 5.955 & 70 & 3.551 & 0.4017 & 4.235 \\
877.02 & 12.04 & 35 & 5.336 & 0.3169 &   \\
878.01 & 23.59 & 17 & 4.062 & 0.2899 &   \\
880.01 & 26.442 & 16 & 4.722 & 0.2581 & 1.727 \\
880.02 & 51.529 & 9 & 4.583 & 0.1748 &   \\
880.03 & 5.902 & 75 & 3.614 & 0.4038 & 5.344 \\
881.01 & 21.023 & 19 & 2.842 & 0.361 &   \\
882.01 & 1.957 & 225 & 8.348 & 0.3828 &   \\
883.01 & 2.689 & 158 & 5.226 & 0.4033 &   \\
884.01 & 9.439 & 45 & 4.717 & 0.3587 & 4.222 \\
884.02 & 20.473 & 18 & 9.076 & 0.0853 &   \\
886.01 & 8.012 & 56 & 12.379 & 0.246 &   \\
887.01 & 7.411 & 59 & 4.882 & 0.3659 &   \\
889.01 & 8.885 & 51 & 5.279 & 0.354 &   \\
890.01 & 8.099 & 56 & 3.843 & 0.3911 &   \\
891.01 & 10.007 & 43 & 4.104 & 0.3743 &   \\
892.01 & 10.372 & 38 & 3.328 & 0.3875 &   \\
893.01 & 4.408 & 77 & 4.039 & 0.4019 &   \\
895.01 & 4.409 & 102 & 6.172 & 0.3764 &   \\
896.01 & 16.24 & 25 & 5.135 & 0.2931 &   \\
896.02 & 6.308 & 69 & 3.986 & 0.3978 & 2.163 \\
897.01 & 2.052 & 206 & 6.095 & 0.4043 &   \\
898.01 & 9.771 & 44 & 5.533 & 0.3318 & 7.133 \\
898.02 & 5.17 & 83 & 5.88 & 0.3715 & 9.014 \\
898.03 & 20.09 & 22 & 3.83 & 0.32 &   \\
899.01 & 7.114 & 60 & 5.462 & 0.3544 & 7.366 \\
899.03 & 15.368 & 31 & 4.425 & 0.3338 &   \\
900.01 & 13.81 & 25 & 3.451 & 0.3599 &   \\
901.01 & 12.733 & 34 & 4.129 & 0.3671 &   \\
903.01 & 5.007 & 85 & 4.662 & 0.3905 &   \\
904.02 & 27.945 & 15 & 7.083 & 0.1422 & 30.626 \\
904.03 & 42.157 & 10 & 9.098 & 0.1469 &   \\
905.01 & 5.795 & 75 & 3.586 & 0.4033 &   \\
906.01 & 7.157 & 59 & 3.877 & 0.3923 & 3.376 \\
906.02 & 17.648 & 23 & 4.21 & 0.3204 &   \\
907.01 & 16.514 & 20 & 4.509 & 0.2899 & 3.206 \\
907.02 & 30.133 & 11 & 5.368 & 0.1604 &   \\
908.01 & 4.708 & 95 & 5.033 & 0.3896 &   \\
910.01 & 5.392 & 77 & 4.618 & 0.3911 &   \\
912.01 & 10.848 & 38 & 4.554 & 0.3542 &   \\
913.01 & 4.082 & 107 & 4.875 & 0.4012 &   \\
916.01 & 3.315 & 125 & 5.408 & 0.3961 &   \\
917.01 & 6.72 & 65 & 4.416 & 0.3812 &   \\
918.01 & 39.644 & 10 & 11.666 & 0.0186 &   \\
920.01 & 21.804 & 19 & 4.231 & 0.301 &   \\
921.01 & 10.282 & 40 & 5.651 & 0.321 & 3.989 \\
921.02 & 18.119 & 22 & 3.86 & 0.3343 &   \\
922.01 & 5.154 & 82 & 4.003 & 0.4009 &   \\
923.01 & 5.743 & 78 & 4.051 & 0.4009 &   \\
924.01 & 39.476 & 12 & 4.861 & 0.192 &   \\
926.01 & 3.166 & 122 & 4.431 & 0.4051 &   \\
929.01 & 6.492 & 55 & 4.966 & 0.3662 &   \\
931.01 & 3.856 & 117 & 4.408 & 0.4069 &   \\
934.01 & 5.827 & 77 & 4.434 & 0.3921 & 1.908 \\
934.03 & 18.747 & 20 & 3.34 & 0.3498 &   \\
935.01 & 20.861 & 23 & 7.642 & 0.1883 & 16.225 \\
935.02 & 42.633 & 9 & 5.199 & 0.1514 &   \\
936.01 & 9.468 & 46 & 4.248 & 0.3758 &   \\
937.01 & 20.835 & 18 & 6.369 & 0.1998 &   \\
938.01 & 9.946 & 43 & 4.39 & 0.3587 &   \\
940.01 & 6.105 & 75 & 5.239 & 0.3776 &   \\
941.01 & 6.582 & 70 & 3.852 & 0.3997 & 4.005 \\
941.03 & 24.665 & 19 & 4.453 & 0.2833 &   \\
942.01 & 11.515 & 37 & 3.922 & 0.3704 &   \\
943.01 & 3.601 & 121 & 4.316 & 0.4055 &   \\
944.01 & 3.108 & 147 & 5.242 & 0.4043 &   \\
945.01 & 25.847 & 15 & 4.27 & 0.2943 & 5.259 \\
945.02 & 40.718 & 10 & 6.671 & 0.1056 &   \\
947.01 & 28.599 & 14 & 2.722 & 0.336 &   \\
949.01 & 12.533 & 31 & 4.489 & 0.3554 &   \\
951.01 & 13.197 & 35 & 3.398 & 0.3804 & 1.417 \\
951.02 & 33.653 & 12 & 4.239 & 0.2631 &   \\
952.01 & 5.901 & 53 & 6.798 & 0.3203 & 28.179 \\
952.02 & 8.752 & 41 & 8.306 & 0.2758 & 7.633 \\
952.03 & 22.781 & 14 & 3.532 & 0.2878 &   \\
953.01 & 3.584 & 117 & 4.086 & 0.4098 &   \\
954.01 & 8.115 & 51 & 4.315 & 0.376 & 1.848 \\
954.02 & 36.926 & 10 & 3.942 & 0.2351 &   \\
955.01 & 7.039 & 55 & 4.073 & 0.3847 &   \\
956.01 & 8.361 & 52 & 3.914 & 0.385 &   \\
960.01 & 15.801 & 25 & 6.668 & 0.244 &   \\
961.01 & 1.214 & 336 & 4.783 & 0.4207 &   \\
972.01 & 13.119 & 35 & 4.512 & 0.3461 &   \\
974.01 & 53.507 & 9 & 3.09 & 0.2403 &   \\
975.01 & 2.786 & 148 & 4.643 & 0.4087 &   \\
976.01 & 52.569 & 9 & 6.608 & 0.2361 &   \\
977.01 & 1.354 & 298 & 6.448 & 0.4116 &   \\
984.01 & 4.288 & 103 & 33.905 & 0.0325 &   \\
986.01 & 8.188 & 54 & 3.806 & 0.3884 & 3.806 \\
986.02 & 76.051 & 6 & 9.15 & 0.0079 &   \\
987.01 & 3.179 & 127 & 3.89 & 0.4122 &   \\
988.01 & 10.381 & 40 & 3.491 & 0.3844 & 2.454 \\
988.02 & 24.57 & 18 & 3.458 & 0.3285 &   \\
991.01 & 12.062 & 34 & 3.604 & 0.3804 &   \\
993.01 & 21.853 & 21 & 4.709 & 0.2912 &   \\
999.01 & 16.568 & 27 & 5.125 & 0.3164 &   \\
1001.01 & 40.811 & 11 & 4.576 & 0.1958 &   \\
1003.01 & 8.361 & 49 & 10.766 & 0.2186 &   \\
1005.01 & 35.618 & 13 & 3.289 & 0.3114 &   \\
1014.01 & 17.317 & 21 & 3.994 & 0.3121 &   \\
1015.01 & 9.429 & 22 & 3.839 & 0.3257 &   \\
1017.01 & 17.445 & 24 & 4.224 & 0.325 &   \\
1020.01 & 54.356 & 8 & 6.147 & 0.0762 &   \\
1022.01 & 18.827 & 20 & 3.265 & 0.3408 &   \\
1024.01 & 5.748 & 73 & 4.718 & 0.3886 &   \\
1029.01 & 32.309 & 14 & 3.594 & 0.2859 &   \\
1052.01 & 17.028 & 27 & 4.33 & 0.3266 &   \\
1060.01 & 12.11 & 34 & 4.931 & 0.3272 &   \\
1061.01 & 41.815 & 10 & 3.215 & 0.2449 &   \\
1066.01 & 5.715 & 44 & 4.32 & 0.3685 &   \\
1074.01 & 3.771 & 65 & 3.989 & 0.3943 &   \\
1078.01 & 3.354 & 129 & 5.877 & 0.392 & 7.842 \\
1078.02 & 6.877 & 66 & 4.389 & 0.388 & 6.362 \\
1078.03 & 28.463 & 15 & 4.448 & 0.2583 &   \\
1081.01 & 9.956 & 47 & 14.976 & 0.2075 &   \\
1086.01 & 27.666 & 14 & 4.912 & 0.2365 &   \\
1089.02 & 12.218 & 31 & 4.003 & 0.352 &   \\
1094.01 & 6.1 & 69 & 4.694 & 0.3803 &   \\
1095.01 & 51.598 & 8 & 7.2 & 0.0275 &   \\
1102.01 & 12.333 & 37 & 8.927 & 0.2043 &   \\
1102.02 & 8.145 & 51 & 8.892 & 0.2587 & 52.435 \\
1108.01 & 18.925 & 24 & 5.613 & 0.2597 &   \\
1112.01 & 37.808 & 10 & 5.134 & 0.1537 &   \\
1113.01 & 25.935 & 16 & 4.896 & 0.2526 & 1.48 \\
1113.02 & 83.441 & 6 & 2.871 & 0.1906 &   \\
1115.01 & 12.992 & 31 & 5.501 & 0.3062 &   \\
1117.01 & 11.089 & 40 & 5.062 & 0.3357 &   \\
1145.01 & 30.585 & 16 & 9.449 & 0.1592 &   \\
1152.01 & 4.722 & 86 & 3.786 & 0.4089 &   \\
1159.01 & 64.617 & 6 & 3.969 & 0.0786 &   \\
1160.01 & 13.214 & 33 & 5.447 & 0.3057 &   \\
1165.01 & 7.054 & 61 & 4.231 & 0.3902 &   \\
1175.01 & 31.596 & 16 & 4.256 & 0.2801 &   \\
1176.01 & 1.974 & 200 & 5.303 & 0.4106 &   \\
1177.01 & 3.306 & 120 & 7.162 & 0.3734 &   \\
1187.01 & 0.371 & 1160 & 5.402 & 0.4273 &   \\
1191.01 & 8.173 & 19 & 4.732 & 0.2758 &   \\
1194.01 & 8.708 & 15 & 5.682 & 0.1899 &   \\
1198.01 & 16.089 & 28 & 3.486 & 0.3696 & 7.272 \\
1198.03 & 35.678 & 12 & 4.698 & 0.2053 &   \\
1199.01 & 53.528 & 8 & 7.045 & 0.0549 &   \\
1203.01 & 31.882 & 9 & 3.878 & 0.116 & 0.959 \\
1203.03 & 48.656 & 8 & 4.263 & 0.1678 &   \\
1205.01 & 8.639 & 37 & 3.912 & 0.3759 &   \\
1207.01 & 13.735 & 33 & 4.523 & 0.3641 &   \\
1210.01 & 14.554 & 26 & 4.115 & 0.3406 &   \\
1215.01 & 17.324 & 22 & 6.215 & 0.2291 & 5.847 \\
1215.02 & 33.007 & 12 & 6.976 & 0.1325 &   \\
1216.01 & 11.131 & 42 & 5.533 & 0.3294 &   \\
1218.01 & 29.619 & 14 & 3.837 & 0.2793 &   \\
1221.01 & 30.157 & 14 & 5.809 & 0.2662 & 3.767 \\
1221.02 & 51.081 & 9 & 3.814 & 0.2039 &   \\
1227.01 & 2.155 & 187 & 6.104 & 0.3993 &   \\
1236.01 & 35.746 & 13 & 5.468 & 0.1773 &   \\
1236.02 & 12.31 & 34 & 4.721 & 0.3376 & 3.08 \\
1238.01 & 27.072 & 15 & 4.108 & 0.2721 &   \\
1241.01 & 21.406 & 21 & 4.669 & 0.2851 &   \\
1241.02 & 10.501 & 43 & 8.683 & 0.2288 & 10.955 \\
1245.01 & 13.72 & 31 & 3.782 & 0.3595 &   \\
1246.01 & 19.037 & 23 & 6.258 & 0.2474 &   \\
1257.01 & 86.648 & 6 & 8.153 & 0.0192 &   \\
1258.01 & 36.338 & 13 & 4.713 & 0.2223 &   \\
1258.02 & 14.646 & 28 & 4.105 & 0.3426 & 2.7 \\
1264.01 & 14.132 & 34 & 3.556 & 0.3749 &   \\
1266.01 & 11.419 & 38 & 4.268 & 0.3574 &   \\
1270.01 & 5.729 & 74 & 5.271 & 0.3786 & 21.274 \\
1270.02 & 11.608 & 40 & 11.184 & 0.1615 &   \\
1273.01 & 40.058 & 11 & 5.239 & 0.1638 &   \\
1275.01 & 50.285 & 9 & 3.723 & 0.2066 &   \\
1276.01 & 22.789 & 22 & 4.427 & 0.3092 &   \\
1278.01 & 24.806 & 17 & 5.995 & 0.1934 & 5.405 \\
1278.02 & 44.346 & 11 & 3.196 & 0.2982 &   \\
1279.01 & 14.374 & 32 & 3.078 & 0.3858 &   \\
1281.01 & 49.477 & 10 & 4.859 & 0.1586 &   \\
1282.01 & 30.864 & 14 & 4.818 & 0.2211 &   \\
1283.01 & 8.092 & 55 & 3.666 & 0.3937 &   \\
1285.01 & 0.937 & 453 & 11.112 & 0.3897 &   \\
1298.01 & 11.008 & 39 & 4.052 & 0.3713 &   \\
1299.01 & 52.5 & 7 & 5.293 & 0.0874 &   \\
1301.02 & 37.514 & 9 & 6.412 & 0.0577 &   \\
1302.01 & 55.638 & 8 & 4.087 & 0.1501 &   \\
1303.01 & 34.296 & 14 & 3.987 & 0.2766 &   \\
1307.01 & 44.851 & 10 & 5.287 & 0.1574 &   \\
1307.02 & 20.343 & 22 & 3.945 & 0.3219 & 1.296 \\
1308.01 & 23.585 & 15 & 4.952 & 0.2377 &   \\
1309.01 & 10.117 & 44 & 4.615 & 0.3602 &   \\
1310.01 & 19.13 & 21 & 5.573 & 0.3295 &   \\
1311.01 & 83.576 & 6 & 5.116 & 0.0496 &   \\
1314.01 & 8.575 & 41 & 4.94 & 0.3452 &   \\
1315.01 & 6.846 & 64 & 4.406 & 0.3815 &   \\
1320.01 & 10.507 & 15 & 3.552 & 0.2752 &   \\
1323.01 & 3.99 & 41 & 5.003 & 0.3416 &   \\
1325.01 & 10.035 & 37 & 5.234 & 0.3289 &   \\
1329.01 & 33.199 & 12 & 5.657 & 0.1675 &   \\
1332.01 & 19.306 & 13 & 3.248 & 0.2899 &   \\
1336.02 & 15.573 & 23 & 5.858 & 0.246 &   \\
1353.02 & 34.543 & 13 & 3.598 & 0.2544 &   \\
1355.01 & 51.929 & 9 & 5.988 & 0.2621 &   \\
1357.01 & 3.014 & 82 & 4.682 & 0.3899 &   \\
1358.01 & 5.645 & 43 & 4.735 & 0.355 & 8.864 \\
1358.02 & 8.744 & 28 & 4.402 & 0.3326 &   \\
1360.01 & 36.77 & 12 & 3.045 & 0.2988 &   \\
1360.02 & 14.59 & 30 & 5.709 & 0.2903 & 3.984 \\
1361.01 & 59.879 & 8 & 5.967 & 0.0669 &   \\
1362.01 & 20.513 & 12 & 5.73 & 0.1639 &   \\
1364.01 & 20.833 & 17 & 5.057 & 0.2442 &   \\
1366.01 & 19.254 & 25 & 4.485 & 0.3129 & 7.987 \\
1366.02 & 54.157 & 7 & 8.423 & 0.0381 &   \\
1372.01 & 69.668 & 6 & 3.87 & 0.0447 &   \\
1376.01 & 7.139 & 56 & 5.662 & 0.3533 &   \\
1378.01 & 19.302 & 22 & 4.456 & 0.3092 &   \\
1382.01 & 4.202 & 84 & 10.693 & 0.2658 &   \\
1385.01 & 18.61 & 23 & 4.154 & 0.3364 &   \\
1387.01 & 23.8 & 20 & 4.582 & 0.2544 &   \\
1391.01 & 7.981 & 54 & 4.578 & 0.3743 &   \\
1393.01 & 1.695 & 144 & 4.815 & 0.4061 &   \\
1397.01 & 6.247 & 37 & 4.128 & 0.3614 &   \\
1398.01 & 5.823 & 42 & 4.626 & 0.3604 &   \\
1399.01 & 8.75 & 28 & 5.261 & 0.2941 &   \\
1403.01 & 18.754 & 25 & 3.731 & 0.351 &   \\
1406.01 & 11.361 & 38 & 3.158 & 0.3867 &   \\
1408.01 & 14.534 & 32 & 5.05 & 0.3293 &   \\
1409.01 & 16.56 & 25 & 3.655 & 0.3516 &   \\
1410.01 & 15.75 & 28 & 3.523 & 0.3609 &   \\
1412.01 & 37.813 & 12 & 3.337 & 0.2949 &   \\
1413.01 & 12.645 & 34 & 4.003 & 0.3607 &   \\
1419.01 & 1.336 & 259 & 4.771 & 0.4178 &   \\
1420.01 & 13.352 & 17 & 4.848 & 0.2381 &   \\
1422.01 & 5.842 & 61 & 4.296 & 0.3867 & 2.365 \\
1422.02 & 19.85 & 21 & 5.35 & 0.2671 &   \\
1426.01 & 38.873 & 9 & 5.537 & 0.1364 &   \\
1430.01 & 10.475 & 41 & 4.617 & 0.3567 & 2.319 \\
1430.02 & 22.93 & 18 & 5.732 & 0.2174 & 0.995 \\
1430.03 & 77.482 & 6 & 3.702 & 0.1348 &   \\
1433.01 & 19.808 & 19 & 4.692 & 0.2828 &   \\
1435.01 & 40.716 & 11 & 5.543 & 0.1583 &   \\
1435.02 & 10.446 & 44 & 5.548 & 0.3396 & 0.91 \\
1436.02 & 13.751 & 28 & 8.933 & 0.1815 &   \\
1438.01 & 6.911 & 67 & 7.013 & 0.3336 &   \\
1444.01 & 44.932 & 9 & 4.114 & 0.2116 &   \\
1445.01 & 7.169 & 49 & 6.721 & 0.3236 &   \\
1448.01 & 2.487 & 167 & 5.286 & 0.4056 &   \\
1452.01 & 1.152 & 348 & 7.463 & 0.4069 &   \\
1456.01 & 7.887 & 33 & 4.989 & 0.3246 &   \\
1457.01 & 8.028 & 32 & 4.029 & 0.3503 &   \\
1458.01 & 8.981 & 23 & 5.567 & 0.2516 &   \\
1459.01 & 0.692 & 610 & 4.835 & 0.4255 &   \\
1465.01 & 9.771 & 40 & 4.397 & 0.3583 &   \\
1470.01 & 15.434 & 18 & 4.405 & 0.2835 &   \\
1472.01 & 85.351 & 6 & 2.555 & 0.2397 &   \\
1473.01 & 23.02 & 11 & 6.224 & 0.1389 &   \\
1475.02 & 9.513 & 29 & 5.699 & 0.3054 &   \\
1476.01 & 56.362 & 6 & 7.761 & 0.0599 &   \\
1480.01 & 20.381 & 22 & 3.186 & 0.361 &   \\
1481.01 & 5.101 & 50 & 5.347 & 0.35 &   \\
1486.02 & 30.184 & 15 & 3.884 & 0.2661 &   \\
1489.01 & 16.005 & 30 & 5.568 & 0.2906 &   \\
1495.01 & 15.595 & 27 & 3.648 & 0.3482 &   \\
1499.01 & 14.164 & 32 & 5.423 & 0.3165 &   \\
1506.01 & 40.43 & 11 & 4.811 & 0.1765 &   \\
1507.01 & 21.36 & 19 & 3.83 & 0.3056 &   \\
1508.01 & 22.047 & 20 & 4.376 & 0.3018 &   \\
1512.01 & 9.042 & 42 & 5.111 & 0.3347 &   \\
1516.01 & 20.554 & 21 & 3.893 & 0.3256 &   \\
1517.01 & 40.068 & 9 & 3.286 & 0.2464 &   \\
1518.01 & 27.506 & 17 & 5.779 & 0.2097 &   \\
1520.01 & 18.458 & 25 & 4.165 & 0.3283 &   \\
1521.01 & 25.941 & 14 & 3.252 & 0.2637 &   \\
1522.01 & 33.386 & 14 & 4.023 & 0.2744 &   \\
1525.01 & 7.714 & 50 & 3.737 & 0.388 &   \\
1529.01 & 17.976 & 23 & 10.896 & 0.1077 &   \\
1530.01 & 12.985 & 25 & 3.918 & 0.3485 &   \\
1532.01 & 18.115 & 22 & 4.094 & 0.3207 &   \\
1534.01 & 20.422 & 22 & 3.648 & 0.3494 &   \\
1540.01 & 1.208 & 340 & 8.227 & 0.4031 &   \\
1541.01 & 2.379 & 149 & 6.064 & 0.3958 &   \\
1543.01 & 3.964 & 112 & 8.655 & 0.3451 &   \\
1546.01 & 0.918 & 438 & 10.747 & 0.3912 &   \\
1547.01 & 30.694 & 6 & 11.292 & 0.0024 &   \\
1549.01 & 29.481 & 16 & 4.831 & 0.2489 &   \\
1553.01 & 52.759 & 9 & 7.513 & 0.0597 &   \\
1557.01 & 3.296 & 106 & 5.185 & 0.3958 & 4.8 \\
1557.02 & 9.653 & 38 & 4.91 & 0.3405 &   \\
1557.03 & 5.316 & 66 & 4.706 & 0.3811 & 2.261 \\
1560.01 & 31.569 & 14 & 5.218 & 0.2101 &   \\
1561.01 & 9.086 & 26 & 5.659 & 0.2685 &   \\
1563.01 & 5.487 & 42 & 3.348 & 0.3871 & 4.173 \\
1563.02 & 8.291 & 27 & 3.976 & 0.3472 & 3.514 \\
1563.04 & 16.74 & 15 & 5.062 & 0.225 &   \\
1564.01 & 53.449 & 8 & 3.149 & 0.2498 &   \\
1567.01 & 7.24 & 34 & 4.381 & 0.3468 & 7.314 \\
1567.03 & 17.327 & 15 & 5.093 & 0.254 &   \\
1569.01 & 13.752 & 32 & 3.884 & 0.3599 &   \\
1570.01 & 6.339 & 37 & 5.037 & 0.348 &   \\
1572.01 & 9.803 & 24 & 4.678 & 0.3003 &   \\
1573.01 & 24.809 & 17 & 42.306 & 0.0079 &   \\
1576.01 & 10.416 & 41 & 5.558 & 0.3252 & 2.277 \\
1576.02 & 13.084 & 33 & 4.961 & 0.3293 &   \\
1581.01 & 29.547 & 14 & 11.533 & 0.0741 &   \\
1585.01 & 19.179 & 23 & 4.659 & 0.288 &   \\
1587.01 & 52.972 & 8 & 13.883 & 0.0048 &   \\
1588.01 & 3.517 & 127 & 5.206 & 0.3998 &   \\
1589.01 & 8.726 & 52 & 5.428 & 0.3547 & 5.147 \\
1589.02 & 12.882 & 35 & 4.272 & 0.3619 & 4.286 \\
1589.03 & 27.435 & 17 & 3.362 & 0.326 & 3.005 \\
1589.05 & 44.549 & 11 & 4.248 & 0.2159 &   \\
1591.01 & 19.657 & 24 & 3.765 & 0.334 &   \\
1595.01 & 40.109 & 12 & 3.693 & 0.2355 &   \\
1597.01 & 7.797 & 53 & 4.516 & 0.3695 &   \\
1598.01 & 56.475 & 6 & 3.353 & 0.1329 &   \\
1599.01 & 20.419 & 21 & 11.668 & 0.085 &   \\
1601.01 & 10.351 & 37 & 3.666 & 0.3772 &   \\
1608.01 & 9.176 & 45 & 3.984 & 0.3747 &   \\
1609.01 & 41.698 & 10 & 4.577 & 0.2153 &   \\
\enddata
\tablenotetext{a}{This table is published in its entirety in the electronic edition.}
\tablenotetext{b}{\ximax\ values are given for each KOI with its nearest exterior neighbor that satisfies the selection criteria for analysis (i.e., its exterior neighbor that is also reported in this table).}
\end{deluxetable}


\section*{Acknowledgements}
\acknowledgements  Funding for this mission is provided by NASA's Science Mission Directorate.  J.H.S acknowledges support by NASA under grant NNX08AR04G issued through the Kepler Participating Scientist Program.  D. C. F. and J. A. C. acknowledge support for this work was provided by NASA through Hubble Fellowship grants \#HF-51272.01-A and \#HF-51267.01-A awarded by the Space Telescope Science Institute, operated by the Association of Universities for Research in Astronomy, Inc., for NASA, under contract NAS 5-26555.  

\bibliographystyle{plainnat}

\end{document}